\documentclass[smallextended]{svjour3}  
\usepackage{graphicx}

\usepackage{subcaption}
\usepackage[dvipsnames]{xcolor}
\usepackage{booktabs} 
\usepackage[noorphans,vskip=0.5ex]{quoting}
\usepackage{tabularx, makecell}
\usepackage{moreverb}
\usepackage{color}
\usepackage{multirow}
\usepackage{longtable}
\usepackage{listings}
\usepackage{color}
\usepackage{caption}
\usepackage{float}
\usepackage{subcaption}
\usepackage{enumitem}
\definecolor{dkgreen}{rgb}{0,0.6,0}
\definecolor{gray}{rgb}{0.5,0.5,0.5}
\definecolor{mauve}{rgb}{0.58,0,0.82}
\lstdefinestyle{customc}{
  breaklines=true,
  xleftmargin=\parindent,
  language=Java,
  showstringspaces=false,
  keywordstyle=\bfseries\color{green!40!black},
  commentstyle=\itshape\color{purple!40!black},
  identifierstyle=\color{blue},
  stringstyle=\color{orange},
}

\titlerunning{An Exploratory Study of Writing and Revising Explicit Programming Strategies}
\authorrunning{Arab, LaToza, and Ko}

\begin{document}
\title{An Exploratory Study of Writing and Revising Explicit Programming Strategies}
%
%
%
%

\author{Maryam Arab \and
        Thomas D. LaToza \and
        Amy J. Ko
}        

\institute{ Maryam Arab and Thomas D. LaToza \at
    George Mason University, Fairfax, VA, USA \\
    \email{\{marab, tlatoza \}@gmu.edu } \\
        \and
    Amy J. Ko \\
    University of Washington, Seattle, WA, USA \\
    \email{ajko@uw.edu} \\
}

\markboth{In review}%
{An Exploratory Study of Writing and Revising Explicit Programming Strategies}
\maketitle

\begin{abstract}
Knowledge sharing plays a crucial role throughout all software development activities. When programmers learn and share through media like Stack Overflow, GitHub, Meetups, videos, discussion forums, wikis, and blogs, many developers benefit. However, there is one kind of knowledge that developers share far less often: strategic knowledge for how to approach programming problems (e.g., how to debug server-side Python errors, how to resolve a merge conflict, how to evaluate the stability of an API one is considering for adoption). In this paper, we investigate the feasibility of developers articulating and sharing their strategic knowledge, and the use of these strategies to support other developers in their problem solving. We specifically investigate challenges that developers face in articulating strategies in a form in which other developers can use to increase their productivity. To observe this, we simulated a knowledge sharing platform, asking experts to articulate one of their own strategies and then asked a second set of developers to try to use the strategies and provide feedback on the strategies to authors. During the study, we asked both strategy authors and users to reflect on the challenges they faced. In analyzing the strategies authors created, the use of the strategies, the feedback that users provided to authors, and the difficulties that authors faced addressing this feedback, we found that developers can share strategic knowledge, but authoring strategies requires substantial feedback from diverse audiences to be helpful to programmers with varying prior knowledge. Our results also raise challenging questions about how future work should support searching and browsing for strategies that support varying prior knowledge.

\keywords{Programming strategies \and Programming tools \and Debugging \and Social coding}
\end{abstract}

\section{Introduction}
\label{sec:introduction}
An essential feature of modern software development is developers sharing knowledge and expertise with other developers through online communities and direct communication. Developers ask and answer questions on Stack Overflow to educate others \cite{Mamykina:2011:DLF:1978942.1979366,globalSoftwareDevelopment}, and share knowledge through social media, portals, and online chat \cite{Bakhuisen2012KnowledgeSU,panahi,socialMedia}. Open-source software development communities rely on openess for knowledge sharing \cite{Shen2005DevelopingCP}. Developers share knowledge and expertise on social network platforms such as Yammer, Hacker, and others \cite{BeyondAgile,Chau2004KnowledgeSI,Twitter,socialMedia} and in-person by organizing meetups, brown-bags, tech talks, and workshops \cite{BeyondAgile,Chau2004KnowledgeSI,Twitter}, learning from each other inside their organizations and in their broader communities. Across all of these settings, the kind of knowledge that developers share varies immensely, from code to implement specific behaviors,  recommendations of APIs, platforms, and languages, productivity and management tips, and more.

However, one type of knowledge, \textit{programming strategies}, is rarely shared, and appears only to be gained through experience or explicit instruction \cite{Raadt2006ChickSA}. Prior work defines strategic knowledge in programming as any high-level plan for accomplishing a programming task, describing a series of steps or actions to accomplish a goal \cite{latoza2019explicit}. For example, a programming strategy might be a step-by-step process for diagnosing a CSS font defect on a website, a systematic procedure for manually resolving a merge conflict, or a sequence of steps for performing a manual refactoring, among countless other software development activities \cite{DeMillo:1996:CSS:229000.226310,Francel2001TheVO}. Research on program comprehension suggests that ``experts seem to acquire a collection of strategies for performing programming tasks, and these may determine success more than does the programmer's available knowledge" \cite{gilmore1990expert}. Strategies are how developers do their work, and yet how this work is done is rarely shared explicitly.

\begin{figure*}[ht!]
\includegraphics[trim=72 565 72 74, clip, width=1.0\columnwidth, keepaspectratio]{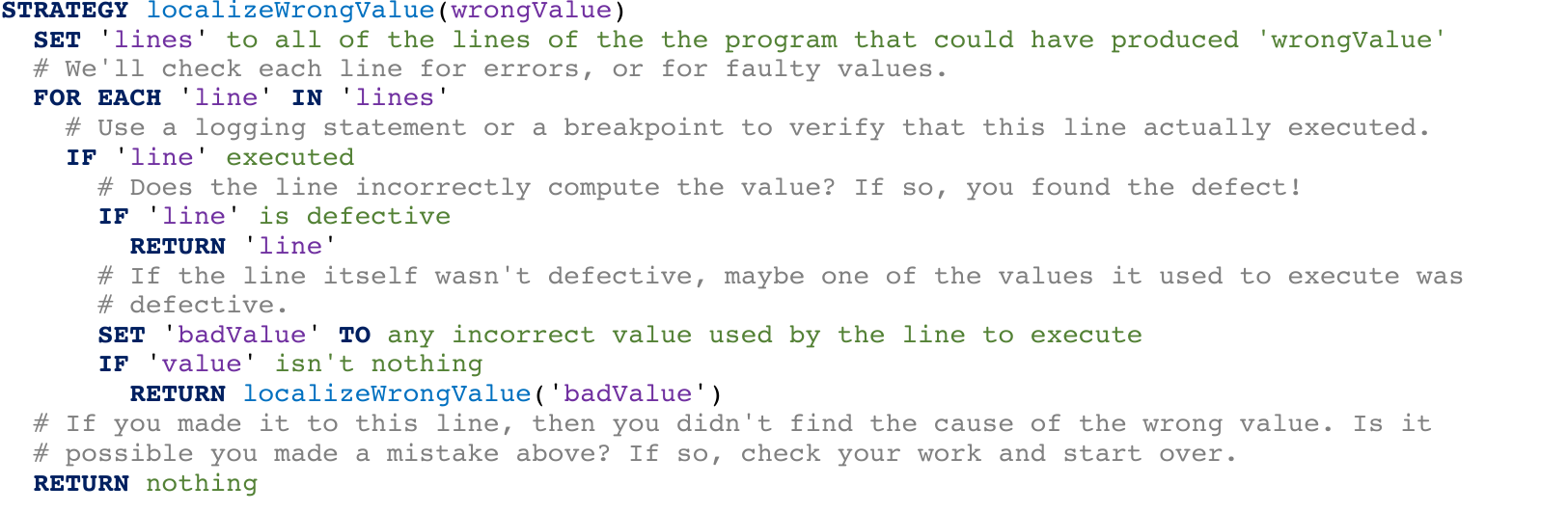} 
\caption{An example strategy for guiding a developers' manual work to localize a line of the code causing a wrong value}	
\label{figure:faultLocalizationStrategy}
\end{figure*}

Creating ways to share programming strategies could reshape software development and computing education. For example, recent prior work shows that when developers are given explicit programming strategies known to be effective (such as the one shown in Figure \ref{figure:faultLocalizationStrategy}, which guides developers in following data dependencies to the source of a wrong value) the effectiveness of their work increases by making them more systematic and efficient \cite{latoza2019explicit}. Similarly, when given explicit debugging strategies, novices are far more successful at debugging when they use explicit strategies than when they use their unsystematic approaches to debugging \cite{ko2019teaching}. Writing effective and explicit programming strategies for the wide range of software development tasks that developers engage every day could unlock the accumulated strategic knowledge of experienced developers, transforming and accelerating software engineering education. It might also even help experts themselves to be more efficient.

Despite the early promise of explicit programming strategies, it is not clear if developers are broadly capable of sharing their strategic knowledge in meaningful, explicit ways. For example, research on knowledge sharing has long shown that some forms of knowledge are \textit{tacit}, in that they are situated, only effectively learned in context, and challenging to articulate explicitly to others \cite{reber1989implicit}. If strategic knowledge in programming is tacit, developers may know how to solve various problems, but only in the moments in which they enact that knowledge, and not in a form that they can recall and share. 

However, it is also possible that strategic knowledge is \textit{not} tacit. Prior work suggests that effective programming is a self-regulated, highly conscious activity \cite{robillard2004effective,loksa2016programming}. This suggests that experienced developers should be able to access their strategies and write them down for sharing. Additionally, software development processes often encode strategic knowledge. For example, methods like test-driven development often provide explicit steps for how to proceed with implementation, which shows with some effort, developers \textit{can} describe the problem-solving processes they use and share them with others.
Finally, research on human expertise shows that while experts certainly develop their own strategies, they also rely on teachers, who provide strategies, and representations for strategies, that allow them to accelerate their learning  \cite{psycologicalExperts}. Therefore, it may be that developers can share knowledge by teaching each other; they simply do not tend to teach strategies in widespread knowledge sharing media.
In this paper, we investigate the possibilities of transferring strategic knowledge to other developers by helping them articulate their strategies. Specifically, we considered:
\begin{itemize}
    \item RQ1: When asked to articulate a programming strategy for particular programming tasks, how do experienced developers' strategies vary?
    \item RQ2: What challenges do experienced developers report with articulating strategies? 
    \item RQ3: When developers make use of explicit strategies articulated by others, what challenges do they face? 
    \item RQ4: When more experienced developers receive feedback from users of their strategies, what challenges do they face in using this feedback to improve their strategies?
\end{itemize}

To investigate these questions, we conducted a study with 34 developers, including 19 experienced developers acting as strategy authors and 15 developers acting as strategy users. Our study simulated a knowledge-sharing platform where authors wrote strategies; users used them on programming tasks and provided feedback. Then the authors attempted to revise their strategies in response to the users' feedback. Throughout, we gathered the strategies that authors wrote, the feedback that users wrote, and the self-reported experiences of all participants in trying to share strategic knowledge by asking them about the challenges they faced by asking survey questions.

In the rest of this paper, we review related background on knowledge sharing in different domains and software engineering (Section \ref{sec:background}) and then describe our study design in detail (Section \ref{sec:method}). We then describe our results, showing that it \textit{is} possible for experts to share their programming strategies in explicit forms. However, they often require substantial feedback to be useful to diverse audiences (Section \ref{sec:results}). We then discuss the implications of this work for software engineering (Section \ref{sec:discussion}) and conclude (Section \ref{sec:conclusion}).
\section{Background}
\label{sec:background}
Prior work on knowledge sharing and strategic knowledge, both in general, and in the domain of software engineering, suggests many potential answers to our research questions. Here, we review this work to help inform our hypotheses and methods.

\subsection{Knowledge Sharing}
Knowledge sharing has long been recognized as a fundamental means by which people contribute to work and innovation in any domain \cite{articleJackson}. There is evidence of achieving more success in educational and industrial environments by sharing strategic knowledge. One study on the relationship of motivational processes to mastery goals found that students who perceived emphasis on mastery goals used more effective strategies, fostering a way of thinking necessary for learning in the classroom \cite{ames1988achievement}. A study examining teaching and learning strategies found that teaching is not only concerned with communicating knowledge but also about developing the skills and strategies for further learning~\cite{McKeachie1985TeachingLS}. An evaluation of teaching strategies in the context of a course showed substantial success in affecting study habits and moderate success for students in affecting achievement later in the semester. 

At the organizational level, work has considered how to foster practical knowledge sharing between employees. Many organizational managers understand the importance of sharing knowledge between employees, and studies have examined the factors impacting an individual's knowledge sharing attitude in the organizational context, and how to align knowledge sharing with organizational attitudes to motivate employees to change their attitude toward knowledge sharing \cite{Bock2001BreakingTM,McDermott2001OvercomingCB}. Other work has identified factors that significantly influence knowledge sharing between individuals in organizations, including the nature of the knowledge, the motivation to share, opportunities to share, and the culture of the work environment \cite{KnowledgeSharingInOrganizations}. One review found that interventions may help create social dynamics within the organization that improves knowledge sharing, including interventions targeting the design of information systems, work structures, and human resource policies~\cite{knowledgeSharingDilemmas}.   
These studies illustrate the importance of knowledge sharing to organizations and the importance of designing effective information systems and interactions between employees to foster knowledge sharing effectively. 

As we noted earlier, research on knowledge sharing has identified two forms of knowledge that can be shared: \textit{tacit} and \textit{explicit} knowledge. Tacit knowledge is implicit, acquired from experience, and constitutes expertise (e.g., as someone learns to ride a bike, they become more capable of riding a bike but less capable of explaining how they do it). The opposite of tacit knowledge is explicit knowledge, which is easily transferable through written or natural language. 
While this basic distinction is persisted in knowledge sharing research, prior work has also found nuanced interactions between the two. For example, knowledge is formed from the interaction of tacit and explicit knowledge \cite{Nonaka}. Similarly, Sanchez \cite{Sanchez} found that tacit knowledge is meaningless without explicit knowledge, so both are complementary and essential for knowledge creation. Other work has found that 'making tacit knowledge explicit' is an important and achievable design goal for IT tools \cite{Polanyi1967TheTD}. While explicit knowledge can be easily articulated and communicated, tacit knowledge is not easily shared \cite{Wang2006KnowledgeSA}, and if it is to be shared, it is costly and slow to externalize \cite{Grant1996TowardAK,Suppiah2009TheIO}. All of these findings, and the theories of knowledge sharing that they support, suggest that programming strategies may indeed be shareable. However, if strategies are largely tacit in nature, doing so will be challenging, and will also require connecting it with other explicit knowledge.

\subsection{Knowledge Sharing in Software Engineering}

Prior work on knowledge sharing in software engineering demonstrates the wide variation in difficulty in sharing different forms of software engineering knowledge. Design patterns are abundant in code, but require effort to organize, describe, and disseminate as knowledge (e.g., ~\cite{Gamma1994DesignPE}). Architectural styles are similarly common, but implicit in code, and have taken non-trivial effort to identify, name, describe, and relate to various software qualities ~\cite{Shaw1996SoftwareA,Bass2012SAIP,Falessi2011DecisionmakingTF}. As software teams generate knowledge, internal tools are often required to help developers externalize, organize, and share it to facilitate collaboration and coordination \cite{Kavitha2011AKM,BeyondAgile}.

In contrast to these more abstract forms of knowledge, the sharing of code, tutorials, and other documents---all of which are explicit in nature---appears to be much easier \cite{socialMedia,Bakhuisen2012KnowledgeSU,Treude2011EffectiveCO,Roehm2012HowDP,Parnin:2011:MAD}. The ubiquity of Stack Overflow \cite{globalSoftwareDevelopment,Mamykina2011DesignLF} and sharing on social media like GitHub, Yammer, Hacker News, blogs, podcasts, and social networks such as LinkedIn, Twitter, Facebook, and Coderwall \cite{BeyondAgile,Chau2004KnowledgeSI,Twitter,socialMedia} reinforce general findings on knowledge sharing: when knowledge is explicit, it will be shared easily and widely.

\subsection{Strategies in Software Engineering}
While there has been abundant prior work on tacit knowledge like patterns and architectural styles, and more explicit knowledge like code, documentation, and tutorials, there has been little work on strategic knowledge in software engineering. 

One idea seemingly related to programming strategies is \textit{programming plans}. Perhaps the most notable work to investigate this was Soloway and Ehrlich's 1984 work, which defined programming plans as ``program fragments that represent stereotypic action sequences in programming'' \cite{soloway1984empirical}. Unlike the programming strategies we investigate in this paper, a \textit{plan} refers to a schema in a programmers' mind for a particular abstract pattern of computation that can be reused for different computational problems. For example, one might imagine a ``search loop plan,'' which is the general approach of using a finite loop and a variable to scan a list to find a single value that satisfies a particular property, such as the maximum value in the list. Numerous works have investigated programming plans, showing, for example, that they are shaped partly by a programming language's affordances \cite{davies1990nature}, that they shape developers' programs independent of language paradigm \cite{green1995programming}, and that learning them is dependent on a robust understanding of a programming language's semantics \cite{xie2019theory}. Therefore, programming plans are fundamentally concerned with algorithm composition and design, and not with procedures for solving the non-algorithm design problems that arise in software engineering (how to test, debug, triage issues, set up branches, etc.).

While plans are orthogonal to strategies, they do interact. For example, prior work has found that programming strategies are a central component of programming expertise \cite{gilmore1990expert,li2015makes,baltes2018towards}. Experts gain strategic knowledge by spending many years creating software and communicating with others.  
Studies have examined the process that expert developers use to solve programming problems, such as listing steps or having phases in a programming task or describing a set of approaches developers choose between. 
For example, one study enumerated a series of steps that developers go through when making decisions about how to reuse code~\cite{Stylos2006MicaAW}.
Other work has described strategies developer choose between when debugging, such as a backwards or forwards reasoning, code comprehension, input manipulation, offline analysis, and intuition~\cite{Bohme:2017:BFE:3106237.3106255}. 
Work has found that researchers, through some effort, can represent programming strategies in an explicit form and teach them to novice developers~\cite{Raadt2006ChickSA,ko2019teaching}.
Studies have also found that having an effective programming strategy can have more of an impact on task success than a programmer's knowledge of plans, design patterns, and other expertise \cite{gilmore1990expert,latoza2019explicit}. Some studies have found that following an effective programming strategy can increase success while reducing task time; for example, a study of slicing strategies for debugging found that those who used slicing had a better understanding of the problem~\cite{Francel2001TheVO}. Another study found that using slicing enhances localizing the fault in the program while debugging~\cite{DeMillo:1996:CSS:229000.226310}. Teaching novice programmers to follow a strategy to systematically trace execution line-by-line and sketch intermediate values can lead to better performance in predicting the output of short programs~\cite{Xie2018AnES}. Developers who follow an explicit programming strategy work in a more organized and structured fashion; strategies enabled their users to finish their tasks more successfully and in fewer time~\cite{latoza2019explicit}.

While researchers studying software developers have been able to explicitly identify and teach new strategies to developers, there has been no prior work investigating the extent to which experienced developers can articulate the strategies they have learned in a form that other developers can use use and follow.
The work we have reviewed here suggests that it can be shared and it is valuable to share; this paper will investigate the challenges faced in sharing.
\section{Methods}
\label{sec:method}
In this paper, we investigate several fundamental questions about the nature of writing and sharing explicit programming strategies and its challenges, the challenges strategy users face, and the difficulties of improving strategies based on feedback from users. To answer these research questions, we conducted a study in which we simulated a knowledge-sharing platform where experienced developers author explicit programming strategies and less experienced developers use them to accelerate their work. Our study consisted of three phases. In the first phase, the first group, which we will call \textit{authors}, each wrote a strategy for a task. In the second phase, the second group, which we will call \textit{users}, each tested two of the authored strategies on two different tasks and provided feedback and comments. In the third phase, each author received the two users' comments and feedback and was asked to elaborate on what might make it challenging to address the users' concerns. 
In the rest of this section, we describe the tasks (Section \ref{sec:Tasks}), the notation participants used for articulating strategies (Section \ref{sec:Roboto}), the participants (Section \ref{sec:Participants}), data collection (Section \ref{sec:Data}),  and the procedure (Section \ref{sec:procedure}).
 
\subsection{Tasks}
\label{sec:Tasks}
\begin{table}
\caption{Authors were each prompted to write a new strategy for one of three programming tasks. 
}
\label{table:authorsTasks-description}
\begin{tabularx}{\textwidth}{X}
\hline\\

\fontsize{8}{10}\selectfont
\textbf{\color{MidnightBlue}Chrome profiler task}

Using the Chrome Profiler require an effective strategy to be successful. For example, a naive profiling strategy might be:
(1) Load the web application,
(2) Start recording performance with the profiler, 
(3) Look at the flame chart, which visualizes how long various components took to render during the recording,
(4) Identify the component taking the most time to render,
(5) Optimize the slow component.
However, this strategy is not particularly effective. It is not intentional about what data to record, whether that recording is representative of a performance problem, and how to make sure sufficient data is gathered to actually diagnose the problem. It also ignores consideration of which parts of the application you may or may not be able to modify to address the issue.

Your task is to write a better profiling strategy. Your goal is to help other developers learn how to effectively profile any web application. Write your strategy in a way that enables other developers to use your strategy and easily identify the components responsible for slow performance. 
Consider the following in writing your strategy:
What steps should a developer follow to accurately identify slow components?
How should they diagnose why those components are slow?
What data should they look at?
Include enough detail so that a developer experienced in web development and JavaScript, but not in profiling, can be successful.
\\\hline\\
\fontsize{8}{10}\selectfont

\textbf{\color{MidnightBlue}Error handling task}

Your task is to write a strategy describing a procedure you use to identify all of the potential errors that might occur in an implemented component and ensure that an appropriate fallback UI message is displayed instead of crashing the component tree. Your strategy should not include steps for implementing error handling logic. It should describe how to identify potential errors and the approach you take to implement error handling for them. 

Write your strategy so that other developers can easily identify where they need to add error handling.
Consider the following in the strategy you write:
What types of errors should the developer consider?
What information should the developer gather as they are investigating potential errors and how these are handled?
What steps should a developer follow to appropriately handle all possible errors which might occur in the component?
Include enough detail so that a developer experienced in web development and JavaScript, but not in writing robust error handling, can be successful.
\\\hline
\fontsize{10}{8}\selectfont
\\
\textbf{\color{MidnightBlue}CSS debugging task}

Your task is to write a strategy describing how you debug a problem in which the visual style of an element is in some way incorrect. Imagine you are working in code written by other developers with which you are not already completely familiar. What steps will you take to find the cause of the incorrect visual appearance? Your strategy should not include steps to resolve the issue, but we need you to write a strategy on the approach you would take to identify the source of fault.

Write your strategy in a way that enables other developers to use your strategy and easily find out how to fix the visual style problem.
Consider the following in the strategy you write:
What issues could result in an element’s visual style appearing incorrectly?
What information should the developer gather as they investigate these issues?
What tools or techniques should the developer use, and how can they help gather the necessary information?
What steps should the developer follow to accurately diagnose the cause of the issue and apply a fix to resolve the issue?
Include enough detail that so that a developer with experience in web development and JavaScript, but not in debugging CSS styles, can be successful.
\\\hline
\end{tabularx}
\end{table}

In selecting tasks, we had several objectives. 
One approach to investigating strategy authoring would be to ask developers to write down a strategy of their choice. However, this might make it challenging to identify strategy users who would need this strategy, who had the appropriate expertise and limits the ability to compare alternative strategies for the same task. Therefore, we selected tasks for which we believed authors could write strategies, and we could identify users and contexts in which they could be applied. We sought tasks that were neither too hard, given the limited time available to participants, or too easy, obviating the need for a strategy. 

We conducted several pilot study sessions with several candidate tasks. We needed the tasks to be familiar for experienced developers to write strategies. At the same time, we needed the approaches for succeeding to be variable enough that we could observe a diversity of strategies. Ultimately we converged on three common front-end web development tasks (Table \ref{table:authorsTasks-description}):

\begin{itemize}
    \item \textit{Chrome profiler task}: write a strategy to use the Chrome profiler to improve a website's performance (first row in Table \ref{table:authorsTasks-description}).
    \item \textit{Error handling task}: write a strategy to verify the robustness of error handling logic in a front-end web application (Second row in Table \ref{table:authorsTasks-description}).
    \item \textit{CSS debugging task}: write a strategy to debug an arbitrary CSS problem on a web page (third row in Table \ref{table:authorsTasks-description}).
\end{itemize}

\noindent In the second phase, strategy users used an explicit strategy on a programming task and reflected on their experiences. Asking developers to use each strategy on authentic tasks of their own would have been ideal. However, finding developers who were encountering these problems in a real context, and when they agreed to participate in the study proved infeasible. Therefore, we instead developed three programming tasks in which strategy users would apply the authors' strategies. In the \textit{Chrome profiler task}, we downloaded a JavaScript codebase with performance issues involving moving images and buttons for adding, removing, and stopping moving images. Adding to the number of moving images caused a significant decrease in the web page's performance. The users were asked to use an explicit strategy to determine the cause of the performance issue and determine how to resolve it. In the \textit{Error handling task}, we developed a JavaScript application containing several error conditions that could occur. Users were asked to use an explicit strategy to identify potential errors that might occur and ensure that each of these error conditions was handled appropriately. In the \textit{CSS debugging task}, we developed a front-end web application with several visual style defects, including an incorrect header color, wrong border color, and incorrect background color for buttons.  The users were asked to find the causes of the defects using one of the explicit strategies.

\subsection{Strategy Description Notation}
\label{sec:Roboto}
There are many forms in which developers might write strategies. We considered an unstructured natural language, a natural language with hierarchical bulleted lists, and other formats. However, to help make strategies as explicit as possible, we suggested a more structured representation used in prior work called Roboto~\cite{latoza2019explicit}. Roboto is primarily natural language but includes simple control flow constructs such as conditionals and loops to enable strategy users be more systematic.
We did not require authors strictly follow the Roboto language in writing their strategies but instead used it as a guideline for how to organize their thoughts and communicate more precisely to strategy users. 

\begin{figure*}[ht!]
\includegraphics[trim=72 370 72 74, clip, width=1.0\columnwidth, keepaspectratio]{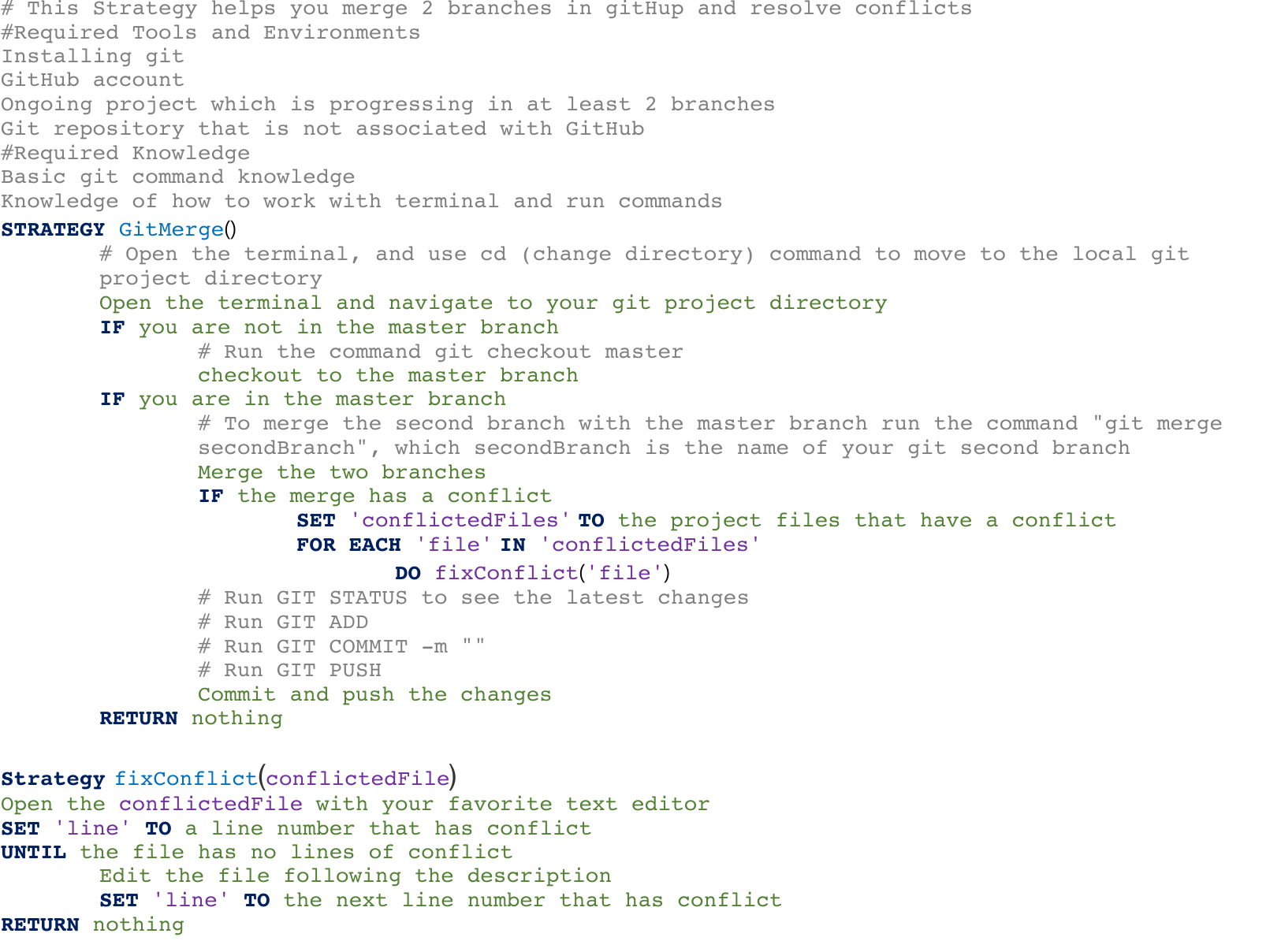} 
\caption{An example of a Roboto strategy for merging two branches in GitHub.}	
\label{figure:mergGithubStrategy}
\end{figure*}

Figure \ref{figure:mergGithubStrategy} lists an example Roboto strategy. It describes how to merge two branches of code in GitHub, including fixing any conflicts. Strategies consist of statements describing what the developer should do next. Statements include performing a specified action, gathering information, or making a decision about how to proceed. Statements in Roboto can be one of six forms: Action, Definition, Conditional, Loop, Return, or Call \cite{latoza2019explicit}. Additional details about each statement can be included through comments, indicated with a hashtag before a statement. 
Strategies may also include preconditions for using the strategy, listed before the strategy declaration. Preconditions describe required knowledge or familiarity the strategy user should have with technologies, resources, programming languages, tools, environments, or platforms.
\subsection{Participants}
\label{sec:Participants}
To recruit developers to author strategies, we sought developers with at least three years of experience in front-end web development in any technology stack. We required author participants to be familiar with one of the three tasks. We recruited participants from developers who were alumni of the authors' institutions. 
 
We asked author participants to evaluate themselves on their expertise with each of the tasks to decide if they were qualified to author a strategy. Three invited authors did not consider themselves qualified and withdrew before we assigned them a task. Nine author participants withdrew after we assigned the task for various reasons, including insufficient familiarity with front-end web development, insufficient experience and confidence to write strategies, not having a procedural strategy, and insufficient time. Of these participants, several expressed interest in participating as user participants. 

To recruit developers to use strategies, we sought developers with a diverse range of programming experience to understand the range of possible difficulties developers with different skills might face in using a strategy. We required user participants to be at least 18 years old and to be familiar with front-end web development, including JavaScript, HTML, and CSS. We recruited user participants from the alumni of authors' institutions and graduate students in computer science and software engineering at both institutions. 

We collected demographic data from both the authors and users. We asked both groups about their prior industrial software development and web-based application development expertise, the number of web-based applications and software application projects they have worked on, and the most significant software or web application they have developed. Finally, we asked them to share a brief qualitative description of their programming and work experience background in a few sentences and a link to any professional profile they might have (e.g., LinkedIn, GitHub).

Our participants included 19 author participants who authored a strategy (identified as A1-A19) and 15 user participants (identified as U1-U15). Table \ref{tab:AuthorsDemographic} gives demographic information for each of the participants. Authors ranged in years of experience in software or web application development from 3 to 48 years (median 9 years) and reported having developed between 0 to 1,000 software applications (median 10) and 2 to 40 web-based applications (median 7). Users ranged in programming experience from 6 months to 9 years with a median of 3 years experience.

\begin{table}
    \centering
    \small
    \begin{tabular}{p{.1\textwidth} p{.04\textwidth} p{.04\textwidth}  p{.04\textwidth} p{.6\textwidth} } 
        \hline
        \centering
        \textbf{Author} & \textbf{SW} & \textbf{Web }& \textbf{Exp } & \textbf{Background}\\\hline
        
        A1&3&7&8& 
        Full stack developer, back-end, ASP.NET, and SQL. \\
        
        A2 & 5 & 2 & 16 & 
        React.js,Vue.js, and other JavaScript frameworks\\
        
        A3 & 14 &10 & 15.5 &
        Systems architect, UX ,full stack developer\\
        
        A4 & 30 & 30 & 20 &
        GitHub, UML, R, Python, C/C++, Java, HTML/CSS, JavaScript, MSON, XAML, C\#\\

        A5 & 20 & 12 & 11 &
        HTML, JavaScript, CSS, PHP,
        JSP, JSF, C\#, Angular\\
        
        A6 & 10 & 5 & 6 &
        Back-end developer, solution design engineer\\
        
        A7 & 0 & 40 & 23 &
        Full-stack web developer, SQL Server and C\#\\
        
        A8 & 10 & 10 & 5 &
        Senior software developer
        \\
        
        A9 & 20 & 20 & 13 &
        MVC with PHP, JS, HTML, and CSS. \\
        
        A10 & 7 & 4 & 5 &
        C\#, 
        JQuery, ASP.NET, 
        CSS, HTML5, and JavaScript 
        \\
        
        A11 & 10 & 10 & 10 &
        Full-stack engineer, multiple large-scale systems\\
        
        A12 & 6 & 5 & 8 &
        Enterprise web applications, SQL in Azure 
        \\
        
        A13 & 30 & 15 & 9 &
        Full-stack web (Python, PHP, JavaScript), network monitoring (Clojure, JavaScript), mobile (Java for Android), VR (C\#, JavaScript), and educational games (Python, JavaScript)\\
        
        A14 & 6 & 3 & 5 & 
        Senior software developer and freelancer\\
        
        A15 & 6 & 6 & 6 &
        Senior software engineer, full-stack development 
        \\
        
        A16 & 10 &5 & 5 &
        Software engineer, open source contributor, freelancer 
        \\
        
        A17 & 50 & 3 & 9 &
        Data processing 
        and data visualization 
        \\
        
        A18 & 6&2 & 3 &
        HTML, CSS, JavaScript, Python, Java, Angular, Node
        \\
        
        A19& 1000 & 15 & 48 &
        My resume last I checked was about 11 pages or so.  
        I have worked at 13 large companies.\\\hline
        
        \textbf{User} & \textbf{SW} & \textbf{Web }& \textbf{Exp } & \textbf{Background}\\\hline
        
        U1&3&3&4& 
        HTML5, CSS, Bootstrap, JQuery, JavaScript, Node.js\\
        
        U2&10&4&3&
        Web developer\\
        
        U3&0&6&1&
        Backend developer, TA for Python and R class 3x\\
        
        U4&4&2&6&
        Junior software developer and student\\
        
        U5&0&3&1&
        Graduate student, HTML, CSS, JavaScript\\
        
        U6&10&7&5&
        Senior software developer\\
        
        U7&3&5&1.5&
        Graduate software engineer student\\
        
        U8&6&2&0.5&
        Software developer student\\
        
        U9&10&4&7&
        Full stack developer, Java, Shell scripting, JavaScript, SQL\\
        
        U10&1&2&1&
        Graduate student\\
        
        U11&0&4&1&
        Graduate student\\
        
        U12&6&2&3&
        Small personal business company\\
        
        U13&6&3&3.5&
        Junior developer\\
        
        U14&7&4&2&
        Software Engineer (full-stack web development, MongoDB, Express.js, ReactJS, and Node.js)\\
        
        U15&3&2&1.5&
        Web developer\\\hline
        
    \end{tabular}
    \caption{Authors and Users' reported development experience:  number of software application developed (SW), number of web-based applications developed (Web), years of experience in software or web application development (Exp), and employment and web technology experience (Background)}
    \label{tab:AuthorsDemographic}
\end{table}
\subsection{Data}
\label{sec:Data}
To answer our research questions, we collected survey responses in each phase of the study in a document. To reduce the bias effect of the proposed questions on the results, we removed the questions from the participants' descriptive responses. We also removed the Likert scale responses to minimize the effects of the proposed questions in the analysis. We prepared three copied documents from the participants' descriptive responses to be analyzed separately by all the paper authors.

In the authoring phase, after writing their strategy, the authors were asked to complete a survey about the difficulties they faced. We brainstormed difficulties they might face when authoring a strategy and prompted them to reflect on these potential difficulties to help them recall their experiences. Authors rated their level of agreement on a 5 point scale with seven potential difficulties and then briefly described their experiences for each difficulty (Figure \ref{figure:authorsSurvey}(a)). Authors were then prompted to share any other difficulties they experienced (Figure \ref{figure:authorsSurvey}(b)). 

In the testing phase, after users finished using each of the two strategies, we asked them to consider five questions about their experiences with the author's strategy. We asked what made it challenging to work with the strategy, and if the challenge was related to a specific line of the strategy, we asked them to specify the related line. Second, the users were asked to consider what aspects or features they thought were missing in the strategy. Third, the users were asked what additional information or details would make it easier for them to follow the strategy. Fourth, we asked them to consider if the strategy was clear and, if not, what made it confusing or ambiguous. Finally, we asked to consider any other challenges they faced while using the strategy.

In the revision phase, we sent the users' comments and feedback to each strategy's author through a shared document. As each strategy was tested with two users, each author received two responses. 
For each user's response, we asked the authors to consider each comment and describe the extent to which the comment makes sense to them, what might make it hard to address, and what aspects might have led them to forget to consider it or makes it hard to consider.

\subsection{Procedure}
\label{sec:procedure}

The study consisted of three phases and was conducted in an asynchronous format entirely remotely through email and dedicated web pages for each phase. We chose to conduct the study asynchronously rather than synchronously for two reasons. 
An asynchronous format better reflects the future context in which we expect strategy sharing to occur, as neither authors or users would be observed. The second reason was the difficulty of scheduling of experienced developers. An asynchronous format enabled them to find time within their schedule or break up their work into smaller time slots as necessary. 

\subsubsection{Phase One:  Authoring}
\begin{figure*}[ht!]
\includegraphics[trim=0 0 0 0, clip, width=1.0\columnwidth, keepaspectratio]{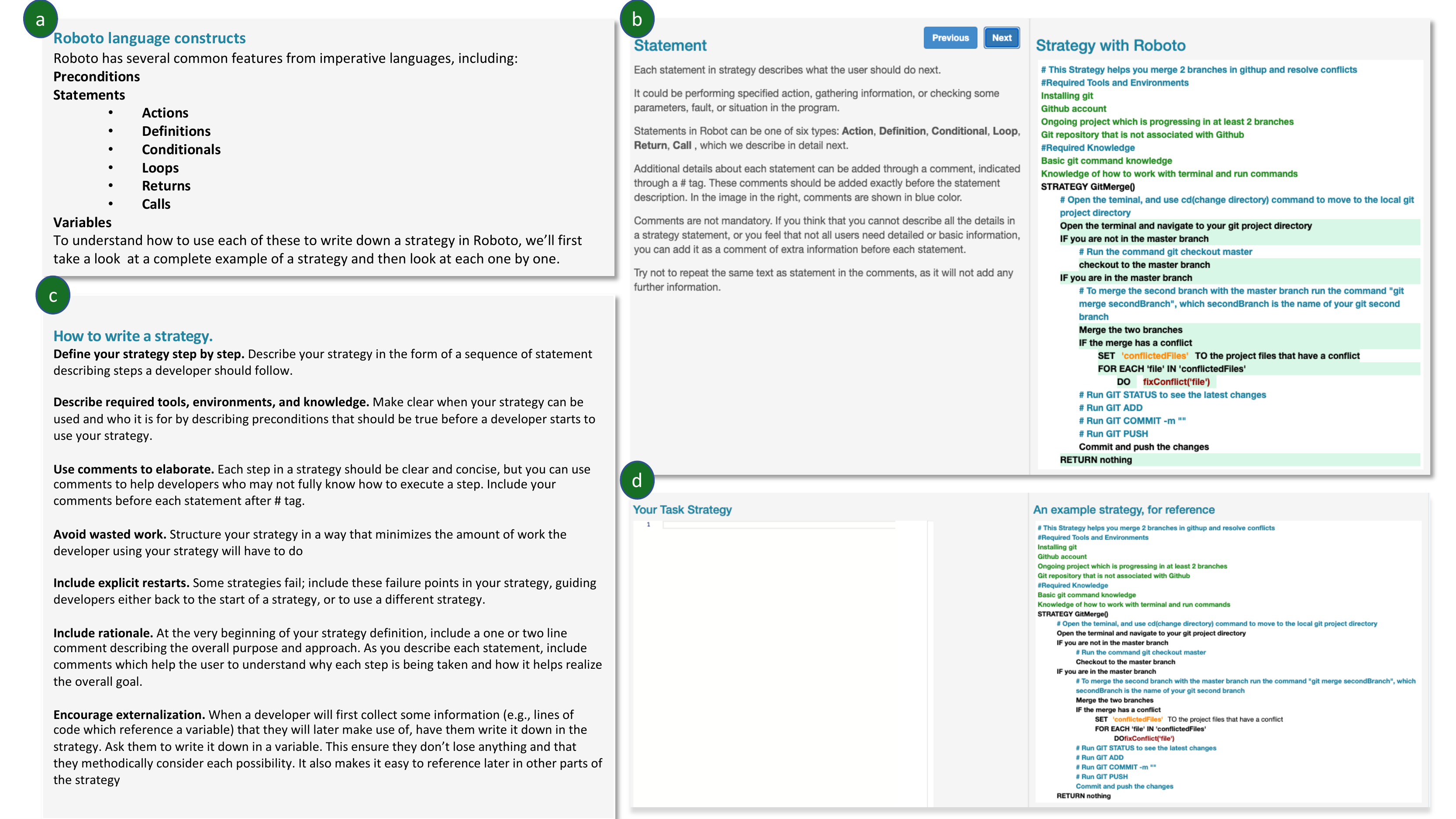} 
\caption{In phase one, authors were given Roboto notation (a,b) and guidelines (c) for writing strategies. Then they are asked to write down one of their own strategies (d) }	
\label{figure:authorsProcedures}
\end{figure*}

After agreeing to participate and selecting one of the three tasks, the authors began phase one by first reading a brief consent form and a tutorial about programming strategies and the syntax of Roboto. Participants were shown an example illustrating a strategy for lifting up state in React and given an overview of the Roboto syntax (Figure \ref{figure:authorsProcedures}(a)).  Participants then completed a tutorial introducing each of the Roboto language constructs (Figure \ref{figure:authorsProcedures}(b)), with a strategy example on the right panel highlighting the related statements for each Roboto language construct in turn. The authors were able to navigate between the tutorial steps to review freely. To help them understand how to write strategies, the authors next read several guidelines for authoring strategies (Figure \ref{figure:authorsProcedures}(c)). The guidelines guided the authors on defining the strategy step by step, describing required tools, environments, and knowledge, using comments to elaborate, avoiding wasted work, including explicit restarts and rationale, and encouraging externalization.  


\begin{figure*}[ht!]
\includegraphics[trim=0 0 105 0, clip, width=1.0\columnwidth, keepaspectratio]{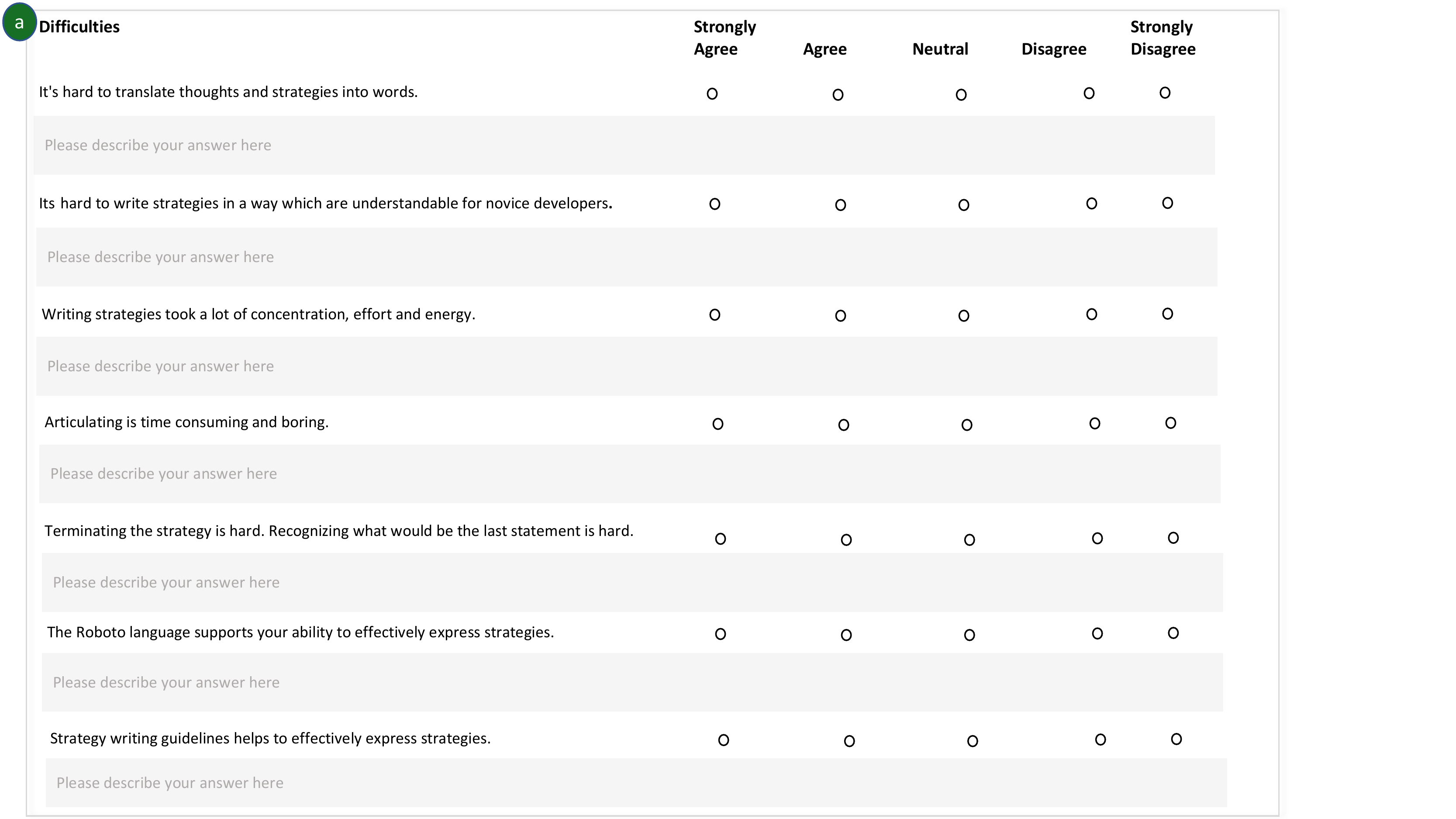}
\includegraphics[trim=0 270 150 0, clip, width=1.0\columnwidth, keepaspectratio]{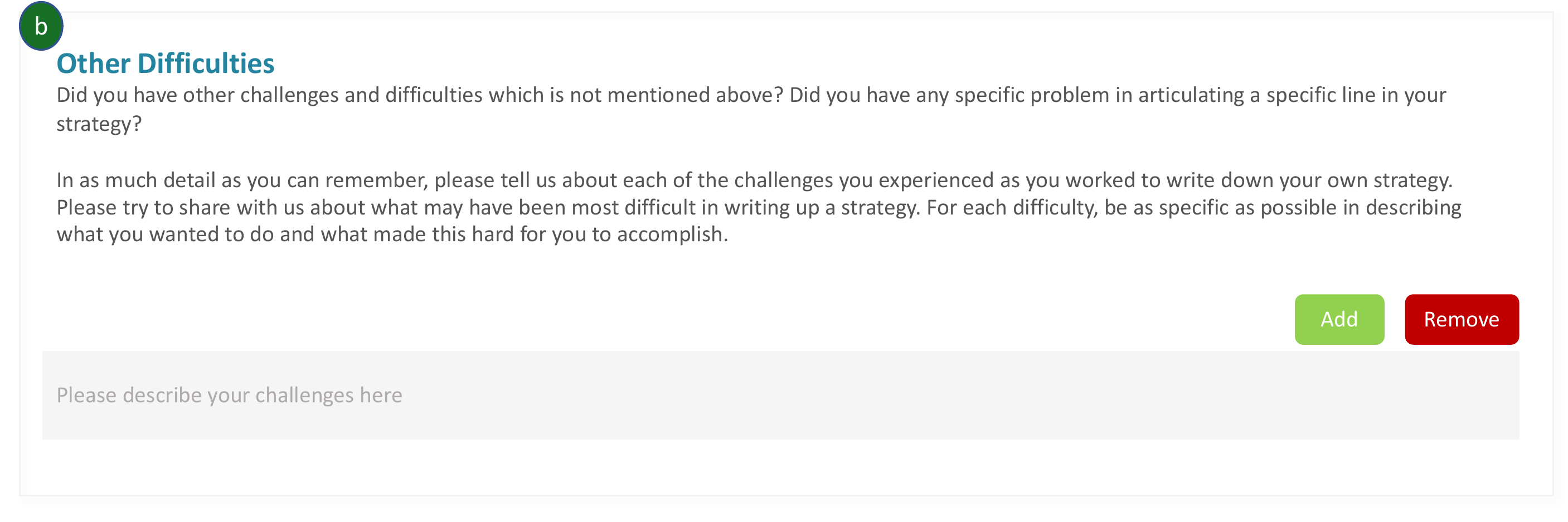}
\caption{In phase one, after the authors finished writing their strategies, they were asked to reflect on the challenges they experienced (a) and share any other categories of difficulties (b). }	
\label{figure:authorsSurvey}
\end{figure*}

The authors next received a task based on their preferred topic of choice, from one of the tasks described in Table \ref{table:authorsTasks-description}, and then they wrote their strategy in a text editor panel (Figure \ref{figure:authorsProcedures}(d)). If authors had difficulties understanding what the task asked them about, they could send an email for clarification. The clarification emails continued until the task became clear to them. Immediately to the right of the panel, the authors could view a sample Roboto strategy. We believed an example would help recall Roboto syntax if they chose to write the strategy in Roboto and the structure we encouraged them to follow. We did not apply any syntax checking or text highlighting in the text editor panel. We meant to give the authors flexibility in selecting their preferred way of describing the strategy in Roboto or natural language. Learning Roboto and concentrating on following its syntax might distract the authors from focusing on their strategy articulation.  The authors then completed the survey on the difficulties they faced (Figure \ref{figure:authorsSurvey}) and completed the demographic items.

The authors were given one week to complete and submit phase one, with a series of reminders and extensions given on days 5, 7, and 12.
\subsubsection{Phase Two: Strategy Use}
\begin{figure*}[ht!]

\includegraphics[page=1, trim=0 220 0 0, clip, width=1.0\columnwidth, keepaspectratio]{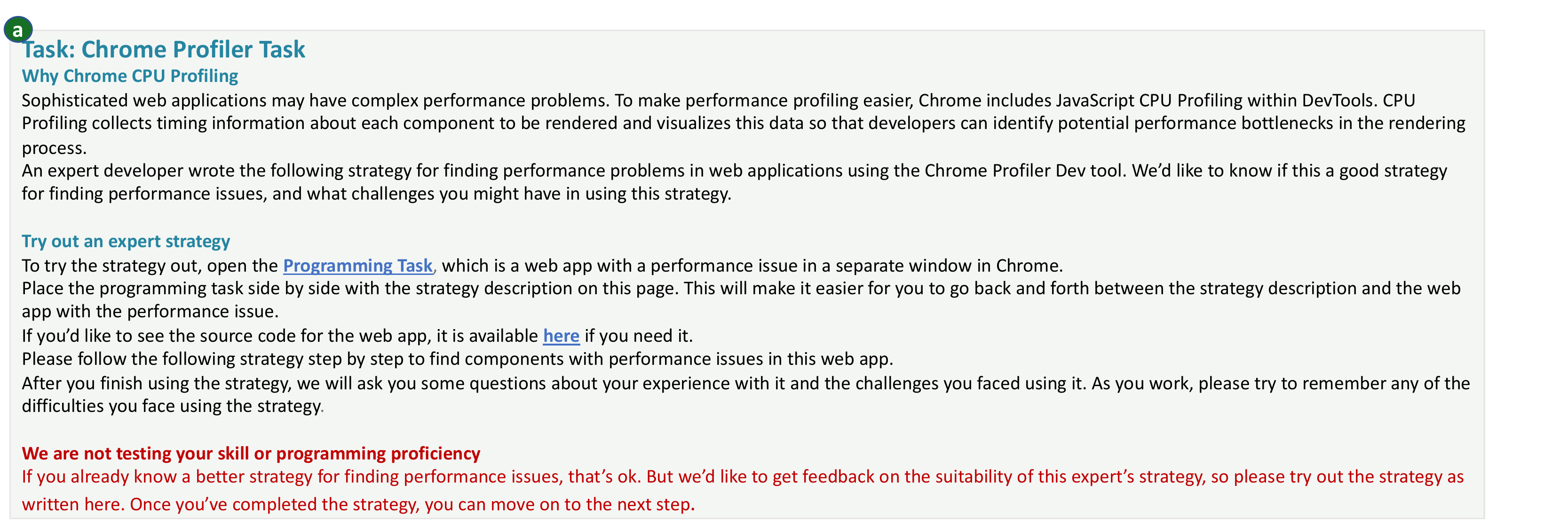} 
\includegraphics[page=2, trim=0 50 0 0, clip, width=1.0\columnwidth, keepaspectratio]{sections/PDF/UserStrategyTesting.pdf} 
\includegraphics[page=3, trim=10 20 0 0, clip, width=1.0\columnwidth, keepaspectratio]{sections/PDF/UserStrategyTesting.pdf}
\caption{In phase two, users were given a programming task (a) and a strategy written by an author (b), were asked to use an online IDE to complete the programming task using the strategy (c), and were then asked about the challenges they faced (d).  }	
\label{figure:testingProcedure}
\end{figure*}


After agreeing to participate, each user began phase two. The users first read a brief consent form and an introduction to programming strategies. They were then given the same detailed tutorial of Roboto language constructs as authors  (Figure \ref{figure:authorsProcedures}(a,b)). They were then asked to try out two authored strategies on two different programming tasks. The users read a task description (Figure \ref{figure:testingProcedure}(a) depicts the description for the Chrome Profiler task) as well as one of the related authors' strategies (Figure \ref{figure:testingProcedure}(b)). To complete the programming task, users received a link in the task description to an online IDE configured with the code (Figure \ref{figure:testingProcedure}(c)). The users were asked to finish the task using the strategy step by step and to try to perform the strategy's actions within the programming task. After finishing, users were asked to complete several survey items about the challenges they faced in using the strategy (Figure \ref{figure:testingProcedure}(d)). After completing the first task, users then applied a second authors' strategy on a second task and again completed the survey items on its challenges. 
Finally, users completed the demographic items. 
The users had four days to complete and submit phase two, with a series of reminders and extensions given on days 3, 4, and 9. If they did not complete the task and survey after day 10, we dropped them from the study.

After each user submitted their feedback on the strategy, one of the experimenters read their feedback. If the feedback was unclear, we asked them followup questions for clarification or to include additional details. We requested the clarification through a shared document with a copy of their responses and the strategy they used, asking the user to clarify marked responses. Multiple rounds of followup communication were performed until we understand the user responses entirely. User participants who completed the study received a \$30 Amazon gift card.

\subsubsection{Phase Three: Revision}
In phase 3, we sent the users' responses about the challenges they faced using the strategy to its author by sharing them through a google document. Each user response was followed by three questions for the author: (1) Does this comment make sense to you; why or why not? (2) What, if anything, makes this comment hard to address? (3) Was there an aspect related to this comment of your strategy, which you forgot to consider; what made this aspect hard to consider? The authors then completed the survey items assessing each response they received. Author participants who completed the study received a  \$40 Amazon gift card.

\begin{figure*}[h!]
\includegraphics[trim=30 10 85 0, clip, width=1.0 \columnwidth, keepaspectratio]{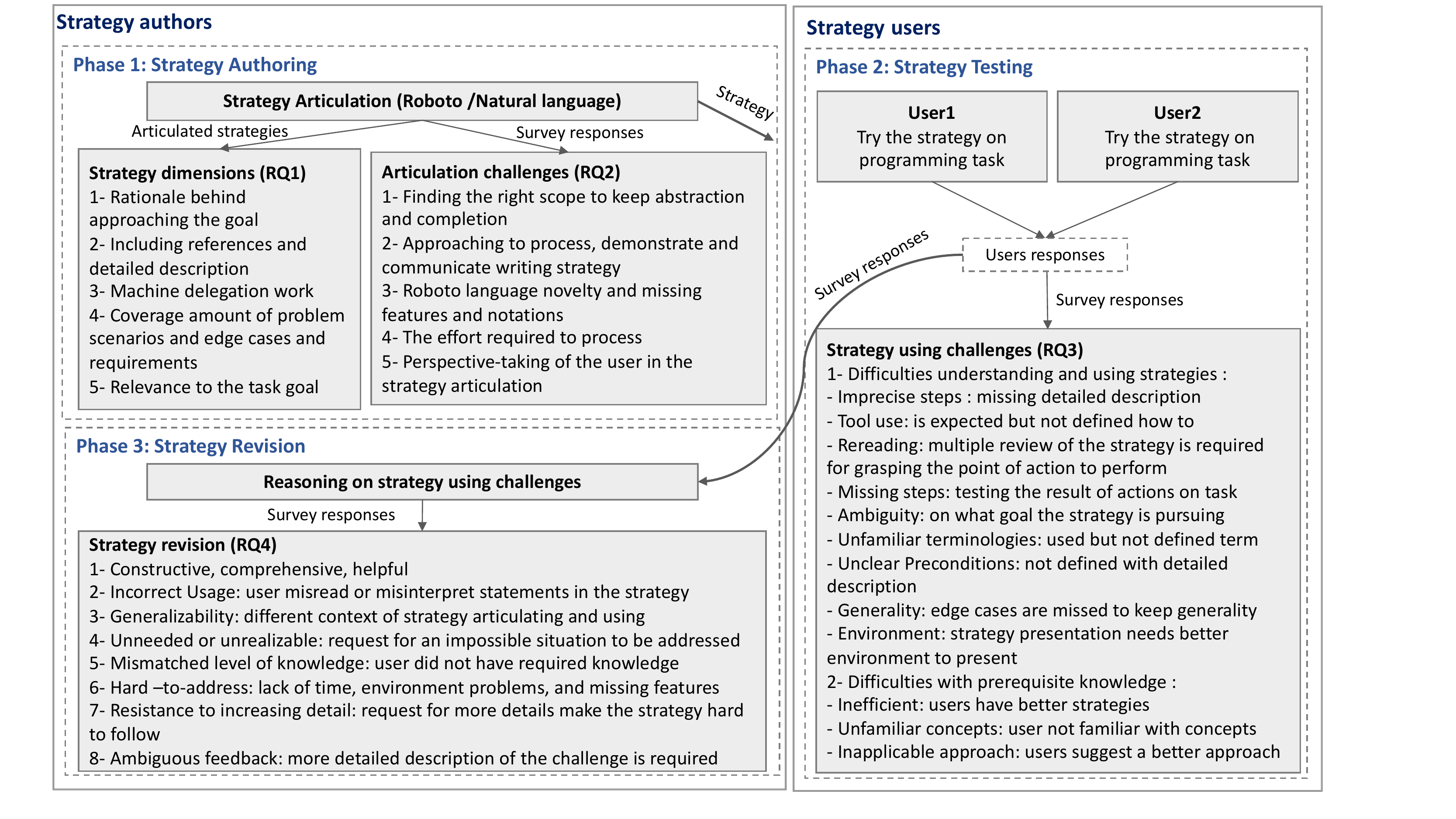} 
\caption{The study procedure and corresponding results.}	
\label{figure:method}
\end{figure*}

\section{Results}
\label{sec:results}
Our analysis focused on answering four research questions. First, we examined how developers choose to articulate and make explicit their strategies (RQ1). Second, as developers articulated their strategies, we examined the difficulties they faced (RQ2). Third, as other developers made use of these strategies on defined related programming tasks, we examined the challenges that strategy users experienced (RQ3). Then, based on the feedback from strategy users to strategy authors, we investigated the possible challenges that strategy authors experienced in improving their strategies based on the feedback they received (RQ4). Finally, we analyzed the relationship between some of the authors' and users' challenges. Figure \ref{figure:method} overviews each phase of the study and how the results were generated. 

\subsection{RQ1: Variation in Authored Strategies}
Overall, we found that all 19 strategy authors were able to write down a strategy for their chosen tasks. These strategies varied in length from 4 lines for a strategy for the Chrome Profiler task to 78 lines for the CSS Debugging task, with a median length of 34 lines. Inspecting each of the strategies, we extracted the sub-goal each group of steps in the strategy was trying to achieve, finding that strategies consisted of four main elements: enumerating potential issues to investigate, determining if the issue applies to the situation at hand, offering a solution plan for addressing the issue, and applying it to edit the code. Some strategies included all of these elements, while others included only some. In the strategy in Figure \ref{figure:coverage-sub}, all four main elements are included.


To characterize how authors chose to express strategies, we identified a set of dimensions along which strategies varied. To do so, in the first round of qualitative coding \cite{Saldaa2009TheCM}, three authors and one collaborator first separately examined all of the strategies to identify inductively identify codes expressing these choices. The codes were then discussed and similar codes merged. The codes were then categorized by authors using pattern coding \cite{Miles1994QualitativeDA}, which identifies smaller thematic sets. The three authors and one collaborator discussed the relationships between codes and merged similar codes. From this, we identified five dimensions, which we discuss below. 

\subsubsection{Rationale}
\label{sec:DimRationale} 
\begin{figure*}[ht!]
\includegraphics[trim=30 590 72 30, clip, width=1.0\columnwidth, keepaspectratio]{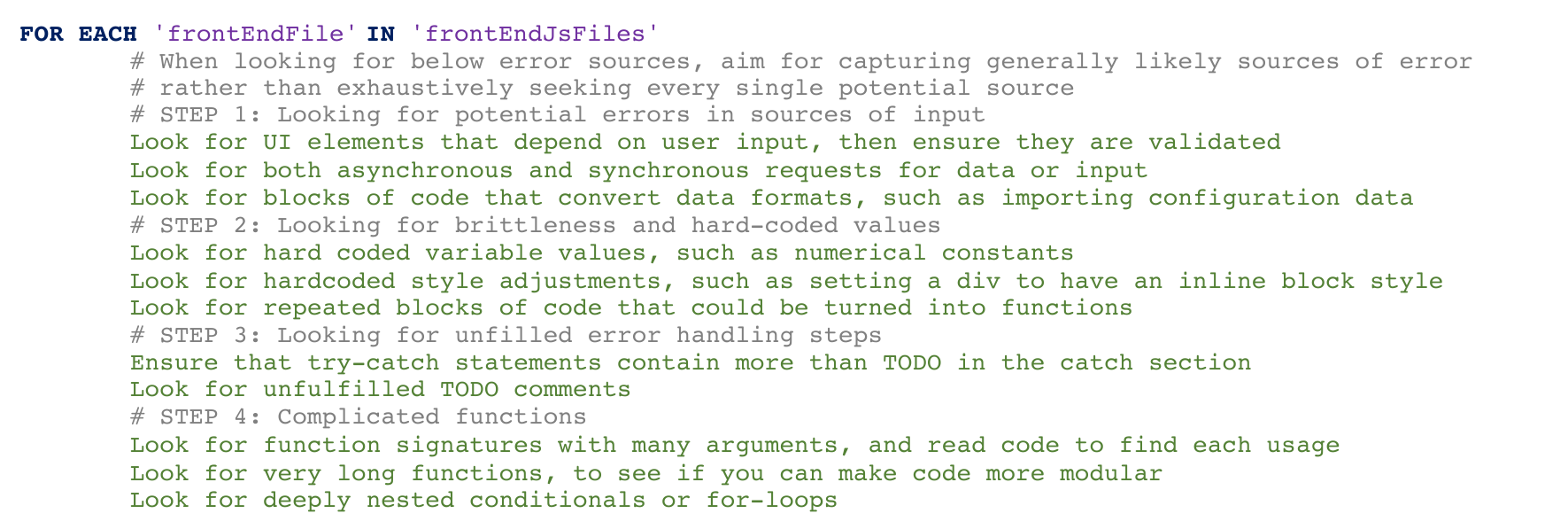} 
\caption{Some strategies included rationale using comments (gray lines beginning with a hashtag).}	
\label{figure:rationale-dimension}
\end{figure*}

\begin{figure}
\begin{subfigure}{1.0\textwidth}
\hspace*{.6in}
\includegraphics[trim=72 620 100 74, clip, width=0.9\columnwidth, keepaspectratio]{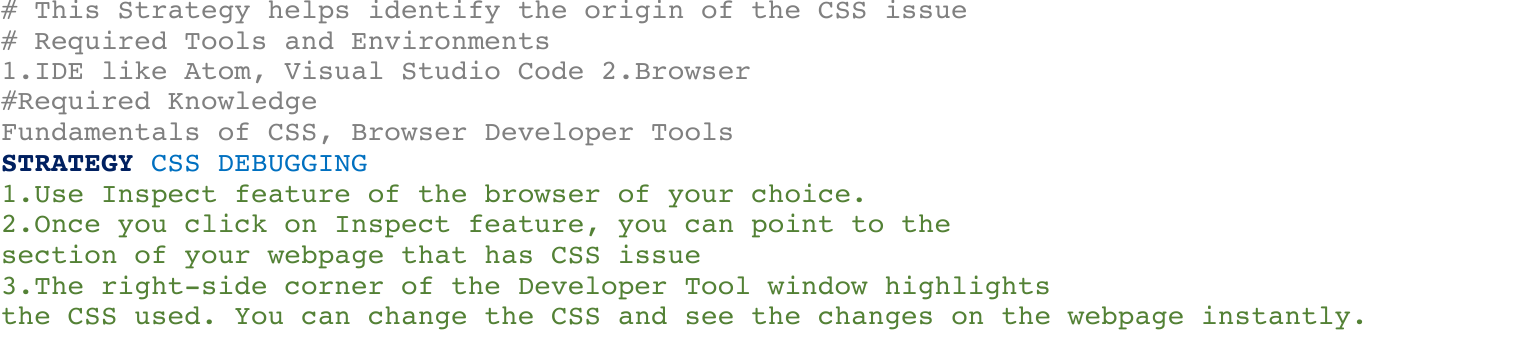} 
\caption{An authored strategy that does not cover exceptional conditions and edge cases} 
\label{figure:no-coverage-sub}
\end{subfigure}
\newline
\newline
\begin{subfigure}{1.0\textwidth}
\includegraphics[trim=0 120 50 50, clip, width=1.0\columnwidth, keepaspectratio]{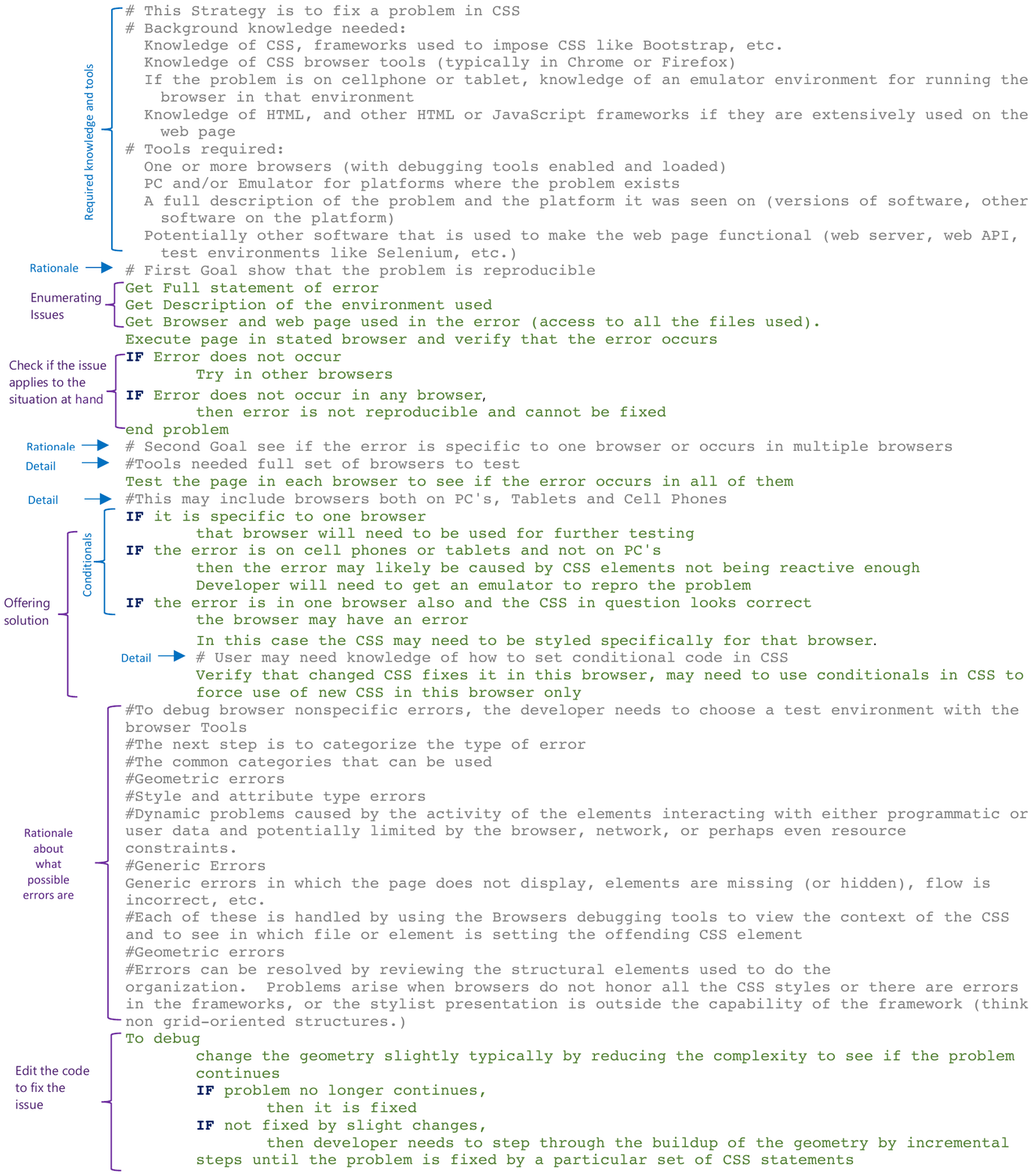} 
\caption{An authored strategy that covers multiple exceptional conditions and edge cases.} 
\label{figure:coverage-sub}
\end{subfigure}
\caption{Two examples of authored strategies for the CSS debugging task.} 
\label{figure:coverage-reference-dimension}
\end{figure}

Strategy authors varied in how much detail they chose to express in describing why each step of the strategy is required. Offering rationale explained why each step was necessary and the larger goal that sequences of steps were attempting to achieve. Some strategies included no explanation of rationale whatsoever, while others included extensive discussion. For example, Figure \ref{figure:rationale-dimension} lists a section of a strategy in which the author included a comment to explain the goal behind each group of steps. 
In contrast, in the strategy in Figure \ref{figure:coverage-reference-dimension}(a), the author chose not to include rationale, but instead only included a high-level description of the steps a user should take.

\subsubsection{References and details}
\label{sec:DetailDimension}
\begin{figure*}[ht!]
\includegraphics[trim=72 580 72 74, clip, width=1.0\columnwidth, keepaspectratio]{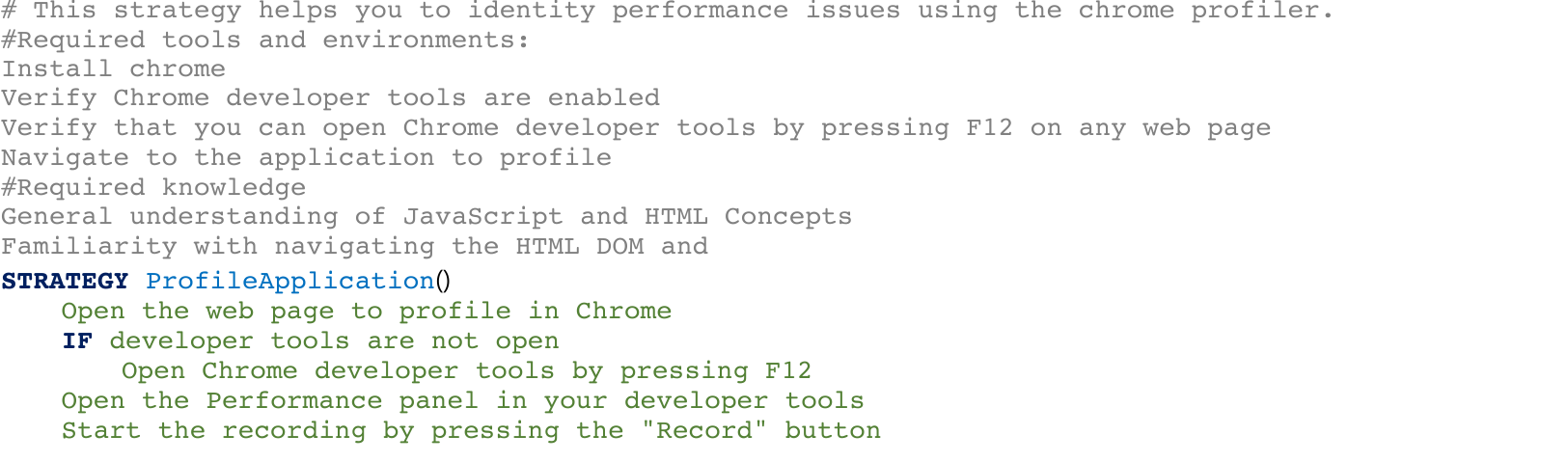}
\caption{An authored strategy which directs the user to set up and check if the Chrome browser and developer tool are ready to be used.}	
\label{figure:setup-dimension}
\end{figure*}

Strategy authors differed in how many references to prerequisite knowledge, tutorials, tool specifications, and tool delegation work they included. Some authors explicitly identified prerequisite knowledge necessary to use the strategy (e.g., the first lines in Figure \ref{figure:coverage-reference-dimension}(b)). Having this knowledge before using the strategy might help users understand the strategy's suitability for their problem. Authors also sometimes described the extent that a strategy is tool-dependent, referencing a specific tool rather than a general category of techniques. Some authors referenced tutorials to provide instruction on how to gain prerequisite knowledge or learn to use the required tools. This included providing links to outside materials and describing how to use the tool to do the strategy's steps. Some authors included additional steps to set up and configure the tools used in the strategy. For example, in the strategy in Figure \ref{figure:setup-dimension}, the author described how the user could make sure that the Chrome Developer Tools are ready to be used for the task.

Authors also varied in the extent to which they elaborated or clarified descriptions of steps for users that might need more detail. Rather than include details in the statements themselves, some used comments. Separating details into comments could help users save time by skipping the detailed description if they do not need it. Some strategies included less detail (e.g., Figure \ref{figure:coverage-reference-dimension}(a)).

\subsubsection{Machine Delegation}
Some authors used Roboto's support for sub-strategies or looping through lists, reducing users' need to monitor their progress using a strategy. Some authors delegated tasks to the machine by defining loops or sub-strategies. The author used loops to record separate scenarios in a list that later used to perform similar actions on each defined scenario.

\begin{figure*}[ht!]
\includegraphics[trim=72 570 72 74, clip, width=1.0\columnwidth, keepaspectratio]{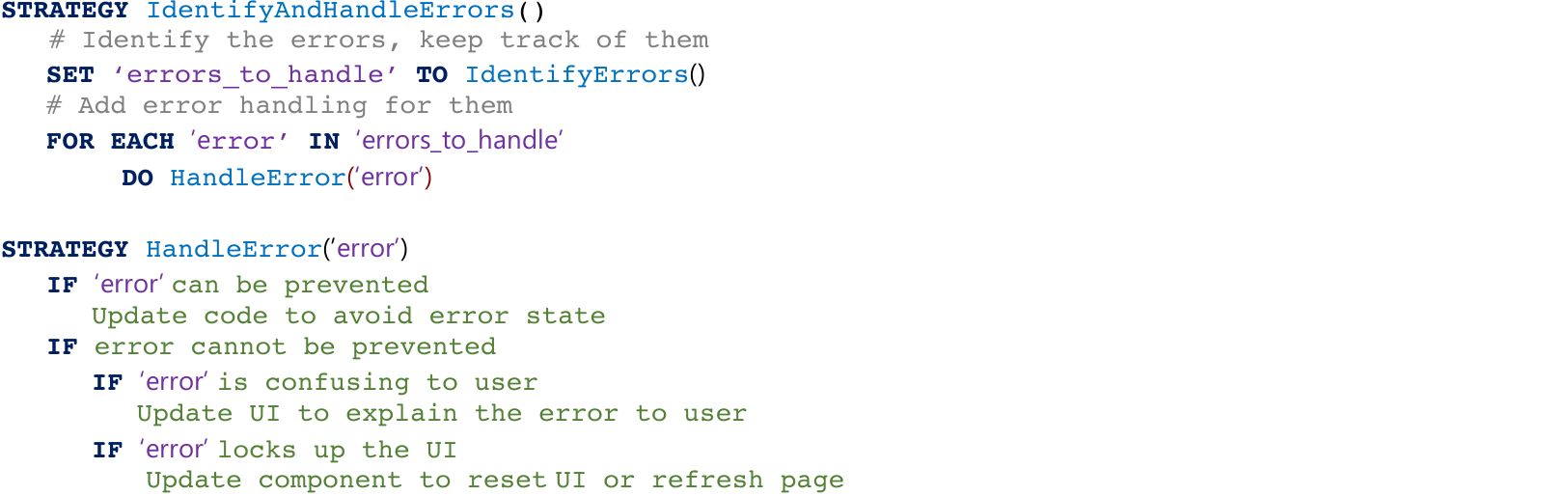} 
\caption{An authored strategy that delegates work to the machine by storing information in a list, `error-to-handle', and iterating over its contents. }	
\label{figure:delegation-dimension}
\end{figure*}

Authors also varied in their use of persisting state by defining variables. Persisting information using state could help users delegate the burden of memorizing required information for future steps they will need to do to the computer. For example, line 3 of the strategy in Figure \ref{figure:delegation-dimension} delegates persistence to the machine by recording information in a variable, `error-to-handle,' and then handling each scenario separately.  
\\

\subsubsection{Coverage}\label{section:Coverage}
Authors varied in their level of generality, abstraction, and coverage of alternative scenarios and edge cases. Strategies differed in the number of different problems the strategy tried to solve, the scope of the problem the strategy covered, and the level of dependency of the strategy to a specific context, language, environment, or platform. Strategies also varied in the number of exceptional conditions they covered. Some considered and checked for many edge cases, while others instead gave a broad overview of the steps that should be taken by the user without considering any exceptional scenarios. Figure \ref{figure:coverage-reference-dimension} shows examples of two strategies written for the CSS debugging task. In the first (a), the strategy considers only one scenario that might occur. In the second (b), the strategy considers several scenarios, using conditionals to consider separate cases. Explicitly testing for exceptional conditionals may help the user to use the strategy more quickly by skipping steps that do not apply.


\subsubsection{Relevance} 
\begin{figure*}[ht!]
\includegraphics[trim=72 530 72 74, clip, width=1.0\columnwidth, keepaspectratio]{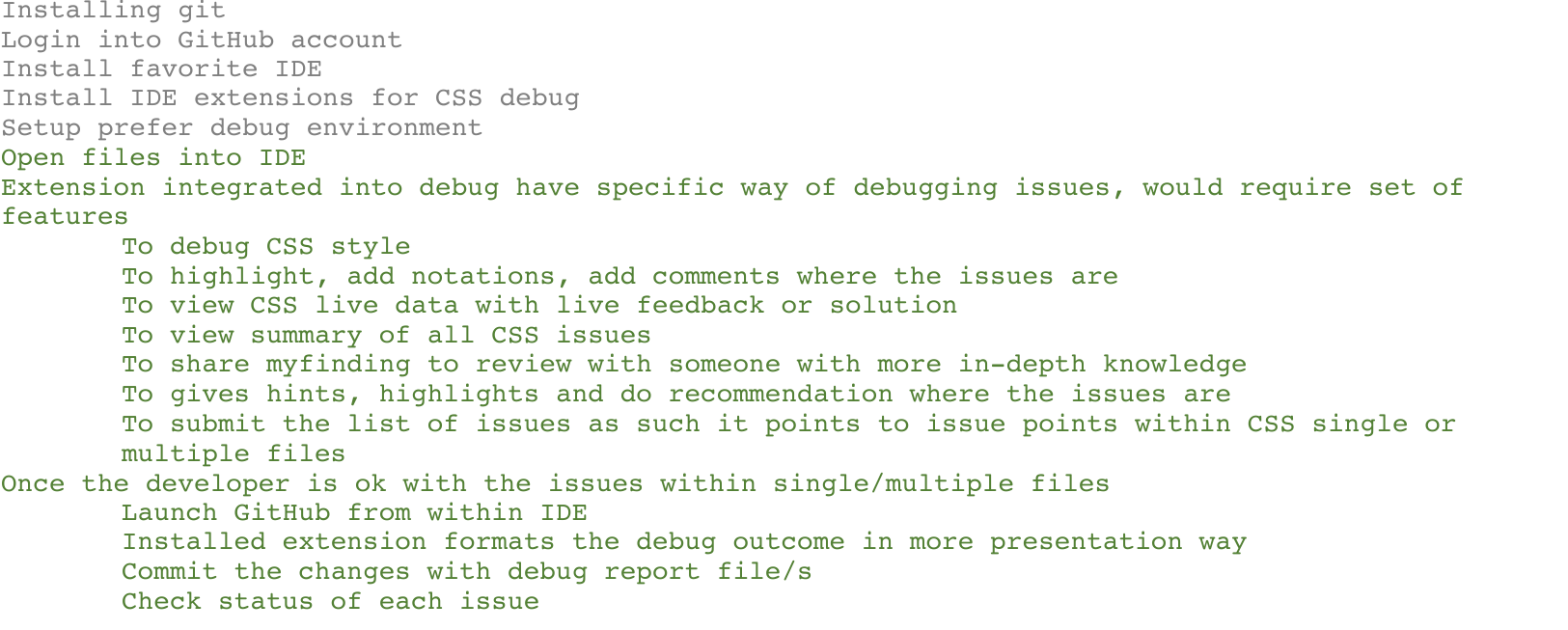} 
\caption{An authored strategy with irrelevant steps.}	
\label{figure:relevance-dimension}
\end{figure*}

Authors differed in their understanding of the goal of the task and the steps they believed to be necessary to include to achieve this goal. For example, one author of the strategy for the CSS debugging task in Figure \ref{figure:relevance-dimension} included steps asking users to install Git, login to a GitHub account, and commit their changes. Interacting with a version control system might well be steps some users might take in some situations when debugging CSS. However, in other situations, users might choose not to commit their changes immediately or use an alternative version control system, making these steps irrelevant to the user's goal.



\subsection{RQ2: Challenges Authoring Strategies}
\label{section:rq2}

\begin{figure*}[ht!]
\centering
\includegraphics[width=0.8\linewidth]{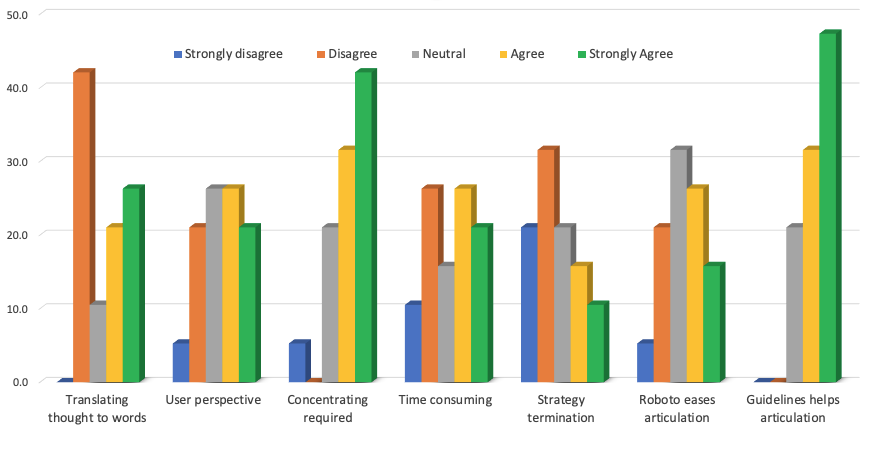}
\caption{Authors' rate (\%) of agreement on the proposed articulation difficulties survey in phase one for 19 authors.}	
\label{figure:AuthorsLikert}
\end{figure*}

To understand the difficulties authors faced in expressing their strategic knowledge, we analyzed free responses given by authors in phase one to prompts to reflect on the challenges they faced in authoring. We received 144 free responses, including responses for each of the 7 prompts for each of the 19 authors as well as 11 responses to the other difficulties prompt. To reduce the leading effect of the prompts given in the survey for proposed difficulties on the data analysis, we reported the questions and the users' Likert selection responses separate from the main analysis. The ordinal-scale agreement survey responses, shown in Figure \ref{figure:AuthorsLikert}, represented that 78.9\% of participants agreed or strongly agreed that having guidelines for articulating strategies helped them to express their strategies adequately and 73.7\% believed that articulation took a lot of concentration, effort, and energy. Interestingly, there were some difficulties for which participants reported different experiences: some strongly agreed that translating their thoughts into words was difficult, others strongly disagreed, with few neutral. The results of the agreement survey are shown in Table \ref{table:authoring-ordinal-difficulties}.

\begin{table}[h]
\small
    \begin{tabular}{cccccccc}
    \hline
    \textbf{} 
    & \textbf{TTW.} 
    & \textbf{UP.} 
    & \textbf{Conc.} 
    & \textbf{Time.} 
    & \textbf{ST.} 
    & \textbf{Roboto.}
    & \textbf{Guide.} 
    \\\hline
(Strongly) Agree & 47.4 & 47.4 & 73.7 &	47.4 & 26.3 & 42.1 & 78.9\\
Neutral &10.5&	26.3&	21.1&	15.8&	21.1&	31.6&	21.1\\
(Strongly) Disagree & 42.1&	26.3&	5.3&	36.8&	52.6&	26.3&	0.0\\\hline
    \end{tabular}
    \caption{Authors rate (\%) of ordinal agreement difficulties report. \\
    \small TTW. = Translating thought to words  UP. = User perspective taking, Conc. = Concentration required, Time. = Time consuming, ST. = Strategy termination, Roboto. = Roboto eases articulation, Guide. = Guidelines helps articulation}
\label{table:authoring-ordinal-difficulties}
\end{table}

To analyze the strategy authors' detailed responses, we created a document of all the data records, excluding the question topic, to prepare for qualitative analysis \cite{Saldaa2009TheCM}. In the first round of qualitative coding, three authors first separately read the responses and inductively identified difficulties as codes with a brief description. Each response was then labeled individually by the three paper authors with zero or more codes. To aggregate these codes, the authors discussed the relationships between codes and merged similar codes. This generated a list of agreed final codes with unique labels for the second round of coding. Using the final codes, the authors then coded the responses a second round. Instances of disagreement were then discussed reach an agreement. This aggregation process iterated multiple times until the authors reached agreement on the coding of all responses. The authors then categorized the final generated codes using pattern coding \cite{Miles1994QualitativeDA}, which groups the similar codes into several broader categories. From this, we identified 25 difficulties across five categories: finding the right scope, approaching writing a strategy, using the Roboto strategy language, the effort required, and taking the user's perspective.

Among the difficulties authors reported, some authors included unsolicited positive feedback about their experiences writing programming strategies. For example:
\begin{quote}
    \small
     \textit{``It helps to communicate problems more clearly. Shows experience, makes us think out of [the] box.''} (A4)
     
    \textit{``It helped me think what I should do for my work project.''}(A10)
    
    
    
    
    \textit{``It is in fact interesting, as it made me realize my process. Many crucial steps are instinctively done and can easily be looked over during the documentation process.''} (A18)
\end{quote}
Much of the positive feedback was about the Roboto language and how it enable articulating strategies to be more structured and easier:
\begin{quote}
\small
    \textit{``Roboto gives it a standardized and structured format which could be easier for any developer to follow. ''} (A1)
    
    \textit{``Roboto language can make it more opinionated to write strategies. It is some kind of standardizing for writing strategies.''} (A8)
    
    
  \textit{``Expressing conditional branching and looping in normal English language can be messy in comparison to Roboto's IF and WHEN statements.''} (A15)
    
  
\end{quote}

In the rest of this subsection, we discuss the five categories of authoring challenges we found.

\begin{table}
\small
    \begin{tabular}{p{2cm} p{3.7cm} p{4.8cm} }
        \hline
        \textbf{Difficulty (frequency)} & \textbf{Description} & \textbf{Authors' quote } \\\hline
        Abstraction (52.6\%)
        &Challenges imagining the range of scenarios to cover in a strategy& 
       \textit{``I had to think about the scenarios I run into and then try to write something which is generic enough without knowing the problem.''} (A1)\\\\
        
        Generalization (57.9\%)& 
        Concerns about strategies being too general to be helpful or too specific to be relevant to many cases& 
        \textit{``It's much easier to write the strategy if the problem statement is more specific, and in a specific programming language.''} (A6)\\\\
        
        Completeness (42.1\%)& 
        Being consistent, structured, and planned, and not forgetting steps that are habitual and tacit&
       \textit{``Many aspects should be taken into considerations. Ignoring or forgetting a small steps may make the whole strategy pointless.''} (A14)\\\\
        
        Scalability (26.3\%)& 
        Scaling to large and more complex problems, which may make strategies hard to understand and use& 
        \textit{``Strategies can end up being too long for complex problems.''} (A14)\\\\
        
        New \newline Information (5.3\%)& 
        Responding to new information during the task& 
        \textit{``Strategy keep[s] evolving based [on] the difficulty of task at hand. ''} (A2)\\\\
        
        Testing (10.5\%)& 
        Ensuring the strategy works well in all cases&
        \textit{``It’s hard to know how one would safely conclude they’ve tested for all possible errors.''} (A11)\\\\\hline
    \end{tabular}\\
  
\caption{Authors reported difficulties in finding the right \textit{scope} of their strategy. The frequency lists the fraction of authors reporting each difficulty.}
\label{table:authoring-general-difficulties}
\end{table}

\subsubsection{Finding the right scope}
The authors reported six challenges related to finding the right scope for their strategy (Table  \ref{table:authoring-general-difficulties}). Some authors found it hard to articulate their strategy in a general way that was suitable for every task and users of varying expertise while keeping the strategy comprehensive and detailed. Authors also found it hard to keep the strategy abstract and imagine the range of scenarios it should cover. Some authors reported that writing a strategy for large and complex problems, testing the strategy, and deciding when a strategy is complete are challenging. Another difficulty was writing strategies in ways that appropriately responded to new information discovered during the task. We discuss two particularly interesting difficulties below.

\textit{1- Generalization:} Some authors found it difficult to write the strategy in a way general enough to be helpful for the many different tasks and situations or felt that their strategy was too specific to be applicable for different tasks. For instance, one author represented: 
\begin{quote}
\small
\textit{``It's hard to describe a strategy for something as abstract as 'UI error handling.' Without a clear description of what the UI is it's hard to know what sort of errors to check for or how.''} (A11)
\end{quote}
Some other authors declared that making a strategy that is generally suitable for any programming task is challenging, while domain specificity reduces these difficulties. As many tools, languages, environments, frameworks, technologies, and aspects may vary across contexts, some believed that specifying a specific setting would simplify writing strategies:

\begin{quote}
\small
\textit{``While it wasn't difficult to write a generic final statement, writing an informative or well-summarizing one seems tricky.''} (A13)
\end{quote}
 
These challenges were ultimately reflected in the differences between the level of coverage between strategies. 

\textit{2- Testing:} Some authors reported that making sure that the strategy would be usable in all possible cases was hard. Some believed that having a program to test the strategy might help create a better strategy by identifying missing steps, conditions, and details. For example, one author reported:

\begin{quote}
\small
   \textit{``Walking through the strategies multiple times to ensure that all broad categories of performance issues and types of slow activities were handled took a decent amount of effort.''} (A15)
 \end{quote}

\subsubsection{How to approach writing a strategy}\label{sec:ApproachDifficulties}

\begin{table}
\small
   \begin{tabular}{p{2cm} p{3.7cm} p{4.8cm} }
        \hline
        \textbf{Difficulty (frequency)} & \textbf{Description} & \textbf{Authors' quote} \\\hline
        Externalizing Strategic Knowledge (15.8\%)
        &Recalling strategies used in the past and translating thoughts into words& 
        \textit{``It’s much easier if it’s done in person or you are talking to that person over writing it down.''} (A2)\\\\
        
        Organization (15.8\%)
        & Need for more strategy examples to learn how to correctly structure the strategy& 
        \textit{``While the guidelines were helpful, I felt the example given was a lot more concrete than the one we were asked to write a strategy for and hence it was limited in how helpful it was.''} (A11)\\\\
       
        Unclear \newline Task (5.3\%)
        &Unclear what the task asked them to write a strategy for&
        \textit{``I almost gave up on this a few times because it seemed so unclear what was being asked of me or how I’d even go about writing such a strategy.''} (A11)\\\\
        
        \multicolumn{1}{l}{Process (26.3\%)}
        & Determining how to effectively frame solving the problem  
        & \textit{``The instructions say not to include instructions on resolving the problem, but some level of resolution is necessary to make sure the change actually worked correctly.''} (A9)\\\\
       
        Usability (10.5\%)
        &Assuring the authored strategy works well with real programs & 
        \textit{``I think it provides a nice framework, but am not sure how well it would work in the real world for specific problems instead of generic concepts.''}(A17)\\\\
        
        Demonstration (21.1\%)
        &Illustrating the strategy without the ability to demonstrate it on a real programming task or supporting tools for communicating necessary concepts& 
       \textit{``In person explanation probably is better by going over debugging strategies using live examples and demonstration''} (A2)\\\\
        
        Choice \newline and Repetition (10.5\%)
        &Articulating reasons for making specific choices between alternative approaches and capturing similar aspects in the strategy to reduce repetition& 
        \textit{``Once I hit a point where there are many directions to investigate it is more difficult to explain my thinking.''} (A12)\\\\
        
        Level of Detail (31.6\%)
        &Finding and expressing the right level of detail to explain strategy adequately& 
        \textit{``It's been a long time since I was a ``novice'' developer and I don't look at a single guide to do anything, I normally piece information from various sources as well as my experience together, so it's hard to know how much detail or not to include.''} (A9)\\\\
        
        \multicolumn{1}{l}{Tool Use} 
        \newline (21.1\%)
        &Communicating necessary terminology and concepts for using referenced external tools& 
        \textit{``It's easy to look at the Profiler and know what's happening but it's hard to explain it.''} (A3)\\\hline
    \end{tabular}

\caption{Difficulties in how to approach writing a strategy. The frequency lists the fraction of authors reporting each difficulty.}
\label{table:authoring-strategy-approaching-difficulties}
\end{table}

Some of the strategy authors reported difficulties in how to approach writing a strategy. Table  \ref{table:authoring-strategy-approaching-difficulties} enumerates the nine different challenges that we observed. These encompassed challenges with ambiguity in the task, how to articulate implicit knowledge, demonstrating the strategy to users without external resources or aids, effectively frame solving the problem, and expressing the right level of detail in the strategy. 

Two of the challenges were particularly revealing.

\textit{1- Externalizing Strategic Knowledge:} Some authors found it hard to translate their thoughts into words, as it required them to recall strategies used in the past in different problem scenarios:
\begin{quote}
    \small
    \textit{``I just need to remember all the situations I was in and how I resolved the issues.''} (A2)
\end{quote}
Some authors claimed that describing the strategies verbally and elaborating the details to a specific audience is much easier than writing them down:
\begin{quote}
    \small
    \textit{``It is hard because, for many developers, we do not spend time to write; instead, we are focusing more on coding. If we are in a meeting and explain the way we do things will much easier than write them out in a document.''} (A10)
\end{quote}

\textit{2- Level of Detail:} Some authors found it hard to find and express the strategy with the right amount of detail for the level of user expertise. This mirrors the differences between strategies in the level of detail authors chose to include (Section \ref{sec:DetailDimension}). Some authors needed to see more examples of strategies related to varying levels of users' expertise:
\begin{quote}
    \small
    \textit{``Having more examples targeted at various expertise levels would help.''} (A9)
\end{quote}
Some other authors had challenges to keep the correct level of details to make strategy simple at one hand, and on the other hand, easy to understand and follow: 
\begin{quote}
    \small
    \textit{``To make the description easy to follow and understand, I'd probably be leaving out a lot of edge cases and essential information.''} (A11)
\end{quote}
\begin{table}
    \centering
    \small
     \begin{tabular}{p{2cm} p{3.7cm} p{4.8cm} }
        \hline
        \textbf{Difficulty (frequency)} & \textbf{Description} & \textbf{Authors' quote} \\\hline
        
        Expressiveness (42.1\%)
        &Missing language constructs required to express strategy in way that is clear and concise& 
        \textit{``I wish there was more coded syntax. It was a bit too much like pseudo-code in that I felt I can make up syntax. ''} (A17)\\\\
        
        Learning novel language (5.3\%)
        &Novelty of using Roboto& 
        \textit{``Using Roboto is new. I need to get used to it.''} (A7)\\\\
        
        Formal \newline Notation (10.5\%)
        &Expressing ideas that are simple to say in natural language more formally in Roboto syntax& 
        \textit{``It gives it a standardized and structured format which could be easier for any developer to follow. They need to be familiar with some programming style though. 
        For example, a developer who mainly works with HTML and CSS may not quickly grasp the syntax and format mentioned for Roboto strategy. ''} (A1)\\\\
        
        Authoring Tools (5.3\%)
        &Missing support in the strategy editor for syntax highlighting, code formatting, signalling line breaks, toolkits& 
        \textit{``I wish there was more coded syntax. It was a bit too much like pseudo-code in that I felt I can make up syntax.''} (A17)\\    \hline
    \end{tabular}

    \caption{Difficulties using the Roboto language and its percent of reports from 19 authors.}
    \label{tab:Roboto-related-difficulties}
\end{table}
\subsubsection{Using the Roboto strategy language}
Although we did not make it mandatory to use the Roboto syntax, most of the authors tried to use the Roboto as a syntax to express their strategy, among whom some authors found it difficult (Table \ref{tab:Roboto-related-difficulties}). Some reported that the novelty of the Roboto language made it hard to use. Others found it hard to express clearly in Roboto what they can express easily in natural language. Features that authors felt to be missing in the strategy editor made it hard to use for others.

\subsubsection{Effort required}
Some authors found it hard to write strategies due to the time and effort required (Table \ref{table:authoring-time-difficulties}). Some reported that writing a strategy down was more time consuming than verbal communication. Others reported that it required intense focus and concentration. Some felt that this work was inherently boring, while other authors reported finding it exciting. Authors reported that strategy authoring, similar to other programming language skills that have a learning curve, requires time and effort to learn.
\begin{table}
\small
  \begin{tabular}{p{2cm} p{3.5cm} p{4.5cm} p{1cm}} 
    \hline
       \textbf{Difficulty (frequency)} &\textbf{Description}&\textbf{ Authors' quote} \\\hline
        Time (52.6\%)
        &More time consuming than verbal communication& 
        \textit{``It takes time to write something down in a proper, structured format over just talking to someone or even writing it down based on the exact issue. ''} (A1)\\\\
        
        Concentration (15.8\%)
        & Cognitively demanding task which requires freedom from distraction and could be frustrating and mental exhaustive& 
        \textit{``It needs an understanding of the problems. Asking 3 core questions in any strategies. Why What and How.
        Concentration to keep the key points. The effort to make sure all the use cases are addresses and energy to keep the team going.''} (A4)\\\\
        
        Not Interesting (5.3\%)
        & Not viewed as a fulfilling and enjoyable task& 
        \textit{``I wouldn't want to do this for my job.''} (A3)\\     \hline
    \end{tabular}

     \caption{Difficulties concerning the effort required to write a strategy. 
     }
     \label{table:authoring-time-difficulties}
\end{table}

\subsubsection{Perspective taking}
About 63\% of the authors reported challenges with perspective-taking. Some authors found it hard to pick an expected level of knowledge for the strategy's user and ignore their own level of knowledge:
\begin{quote}
\small
    \textit{``It's very difficult to know if somebody else would understand the instructions.''} (A9)
\end{quote}
Some other authors reported difficulty guessing the first question a user might have in using their strategy and including an appropriate description. The authors believed that the proper strategy is very dependent on the user's level of experience:
\begin{quote}
\small
\textit{``Some problems/programming scenarios have multiple layers of complexity. To write strategy for novices, these complexities must be abstracted but it should also make sense. Achieving this balance is difficult. ''} (A18)\\
\end{quote}

\subsection{RQ3: Challenges Using Strategies}
\label{section:rq3}
To characterize the difficulties developers face in \textit{using} authored strategies, we analyzed the 150 comments we received from users. 
Using qualitative coding \cite{Saldaa2009TheCM}, three authors separately examined the users' responses, and inductively identified codes with a brief description for each response to represent the user's difficulty in using a strategy. To aggregate the codes, the generated codes were combined, with three authors discussing the relationships between codes and merging similar codes with a unique label to generate a list of agreed final codes. In the second round of coding, the authors labeled the records with the final codes. Then the authors discussed the disagreements, and this process continued until reaching agreement. Finally, the generated codes were clustered into groups using pattern coding \cite{Miles1994QualitativeDA}. This process yielded two broad categories of difficulties: using strategies related to strategy concepts and difficulties related to the user's level of knowledge.

\subsubsection{Difficulties understanding and using strategies}
Users reported eight challenges related to understanding and using strategies (Table \ref{table: testing-strategy-related}). Some found it hard to understand what the strategy asked them to do and the relationship of this to the overall goal. Some users needed a more precise description of how to take a step in the strategy. Some reported that they needed to read the strategy multiple times to understand what it asked them to do. Other users had challenges understanding if the strategy would work for the programming task they had in hand, as the strategy did not support covering relevant edge cases. 

Some of these challenges were directly related to challenges that the authors reported in authoring:

\begin{table}
\small
    \begin{tabular}{p{2cm} p{3.7cm} p{4.8cm}}
        \hline
        \textbf{Difficulty (frequency)} & \textbf{Description }& \textbf{User's Quote} \\\hline
        Imprecise Steps (40\%)&
        Need for more detailed description of some steps. &
        \textit{``Some explanation on how to debug edge cases could make it easier to follow, like how to debug animated elements or how to debug when element's style is manipulated with js. Also, debugging on inline styling could be helpful. ''} (U6)\\\\
        
        Tool \newline Use (13.3\%) &
        Determining how to use a tool to perform the described action, particularly finding a specific referenced functionality in the interface or understanding how to use it&
        \textit{``Like the simple instruction given for opening the chrome developer tool (pressing F12), I was looking for something to teach me how to use the recorded files in order to find and fix the issue.''} (U10)\\\\
        
        Rereading (6.7\%)&
        Need to read strategy multiple times to understand &
        \textit{``I did not know how to use Performance profiler so I had difficulties figuring out the current frame state, network requests, animations but when I followed it twice, I was able to get to it. ''} (U1)\\\\
        
        Missing \newline Steps (66.7\%)
        &Missing instructions or steps to solve the problem.&
        \textit{``The following is missing 1. Make changes and see if they are appropriate. 2. Take those changes and make it back in the IDE.''} (U1)\\\\
        
        Ambiguity (46.7\%)
        &Difficulty understanding the rationale of why a step is required to reach the goal&
        \textit{``In the expert strategy section, there are two points [not] explained. Its not clear are they related to each other or both are different things.''} (U8)\\\\
        
        Unfamiliar Terminology (26.7\%)
        &Using unfamiliar words in the strategy without a definition or description&
        \textit{``The strategy is not very clear as it is not very specific in terms of what part of the code to check or how to implement certain strategies such as a "logical flow".''} (U7)\\\\
  
        Unclear \newline Preconditions (6.7\%)
        &Need for more detailed description of what is required to use the strategy&
        \textit{``I think more specific context regarding the preconditions and statements would be better.''} (U5)\\\\
        
        Generality  (33.3\%)
        &Inapplicability to specific contexts, situations, or edge cases&
        \textit{``More description on what part of the code should I check and what are most probable places for errors happening|? Edge cases should be described to cover and not being missed. ''} (U12)\\\\

        Environment (13.3\%)
        & Better environment for reading strategy, including syntax highlighting& 
        \textit{``I believe that color-coding... would be more suitable in order for one to better understand and follow this strategy. ''} (U5)\\\hline
    \end{tabular}
\caption{Strategy users reported challenges understanding and using strategies.
\label{table: testing-strategy-related}
}
\end{table}

\begin{table}
\small
    \begin{tabular}{p{.14\textwidth}  p{.35\textwidth} p{.4\textwidth} } 
        \hline
        \textbf{Difficulty (frequency)} & \textbf{Description }& \textbf{User's quote}\\\hline
        Inefficient (26.7\%)
        &More effective ways to accomplish the strategy goal than that described. &
    
        \textit{``The instructions and details are kinds of too much and also to mention there should be minor changes like chrome developer tools shortcut(ctrl+shift+I), etc.''} (U14)
        \\\\
        Unfamiliar concepts (26.7\%)
        & Lack of familiarity with the concepts used in the strategy
        & \textit{``I was able to understand this strategy and task as I used the similar strategy "hovering over the defected HTML element" using inspect element. It usually helps identifying and resolving the CSS issue.''} (U8)
        \\\\
        Inapplicable approach (26.7\%)
        & Using a strategy that does not address the problem.
        & \textit{``I wasn't able to change the background color to a random color using this strategy, since it's CSS specific and I think that would need to happen in the JavaScript to set the CSS. However, this can help narrow down which style piece needs to be adjusted to set a color like that.''} (U3)\\\hline
    \end{tabular}
\caption{Users faced challenges due to differences between the knowledge assumed by the strategy and their own level of knowledge. The frequency of each is reported as the fraction of users reporting each difficulty.}
\label{tab:testing-user-knowledge-difficulties}
\end{table}

\begin{itemize}
\item \textit{Imprecise Steps}: Some users reported needing more detail to understand how to perform a step in the strategy. For instance, some users reported that they needed more information on what they should refer to in the code and how to accomplish the action. This challenge mirrors the differences between the level of detail authors included (Section \ref{sec:DetailDimension}) and the difficulties authors had in choosing the right level of detail (Section \ref{sec:ApproachDifficulties}). 

\item \textit{Tool Use}: Some of the users found it hard to use a required tool to perform the described action. They mainly had problems finding the described functionality in the tools the strategy instructed them to use. This challenge reflects the difficulties authors experienced in describing the Tool Use actions that they wished the user to take (Section \ref{sec:ApproachDifficulties}).

\item \textit{Ambiguity}: Some users expressed confusion understanding the rationale behind performing actions in the strategy. They reported being confused about what each step tried to accomplish and why it was necessary. This mirrors differences between authors in the amount of rationale they chose to include (Section \ref{sec:DimRationale}). Other users were confused about the terminology that the authors used in the strategy. In other cases, there were errors in the strategy. In one instance, a user got confused by a strategy that invoked a sub-strategy that was never defined.
\end{itemize}

\subsubsection{Difficulties with prerequisite knowledge}
The second group of difficulties reflected a mismatch of the strategy to the knowledge level of the user (Table \ref{tab:testing-user-knowledge-difficulties}). Some users had more experience and knowledge than the author of the strategy and suggested a better strategy or step for accomplishing the same task. In contrast, some users had insufficient experience and felt that the strategy was not descriptive enough and required more description. This challenge mirrors differences in the level of detail included by different authors (Section \ref{sec:DetailDimension}) as well as the challenges authors reported in finding the right level of detail (Section \ref{sec:ApproachDifficulties}). 

\subsection{RQ4: Challenges Revising Strategies}
\label{section:rq4}
After users' comments on their difficulties were returned to the strategy authors, strategy authors were asked to reflect on how feasible the issues would be to fix and what they believed had caused the user's challenges. To understand these issues, three authors separately generated codes for all the authors' responses with a brief description and labeled the responses. Similar to the analysis for the first and second phases, paper authors collated all codes, merged similar codes, and discussed the differences between codes. This process yielded a set of seven strategy revision difficulties. 

Before we describe these difficulties, it is worth noting that some of the authors found some feedback constructive, comprehensive, and helpful. Some authors agreed with the limitations users reported, agreeing that their strategy was not sufficiently comprehensive and should be improved. For example, two authors reported:

\begin{quote}
\small
  \textit{``Yes. It's a fair criticism and a weakness in the strategy of which I was already aware.''} (A5)
  
  \textit{``Yes, this comment makes sense - the reviewer included the instruction in the feedback.''} (A8)
\end{quote}

In other cases, authors felt that the goal and scope of the strategy they wrote were not well defined, leading to a mismatch between what they wrote the strategy to do and what the user expected. They concluded that offering an example of a step or visually presenting the step through an image would offer clarification and make the strategy easier to follow. For instance, author A3 reported: 
\begin{quote}
    \small
    \textit {``I did not consider all the different sections the user would be looking for.  There are a lot of sections in the profile, so it could be challenging and time-consuming to consider ever[y] deviation.''} (A3)
\end{quote} 

In the rest of this subsection, we discuss the seven strategy revision difficulties.

\subsubsection{Incorrect usage} 
While some authors agreed with comments, others disagreed with user comments, viewing them as reflecting a mistake a user made in using the strategy or a misinterpretation of statements in the strategy. Some authors reported wanting a screenshot of users' work to help them understand what the user was doing and where they were getting stuck. Author A14 reacted:
\begin{quote}
\small 
\textit{``Yes, the first part of this comment is not valid because the tester missed reading the comment for all css files. The second part also might have been misunderstood by [the] tester.''} (A14)
\end{quote}

\subsubsection{Generalizability} 
Some authors realized that the context for which they were writing the strategy differed from the context in which the user was using the strategy. This might occur from a tool that did not support the necessary steps. Some authors also understood that the goal and scope of the strategy they wrote were not well defined, leading to a mismatch between what they wrote the strategy to do and what the user expected. The authors found some comments about the user's expectation to have a strategy that works for all possible similar problem scenarios. It was hard for authors to devise a strategy that describes what to do in every possible situation when there were many possible problem scenarios. Finally, some authors felt that they could not provide more details without reducing the generality of their strategy, and the user's request to change the strategy to cover many specific scenarios would cause a loss of generality. One author stated:

\begin{quote}
    \small 
    \textit{``I felt that the original assignment asked me to be generic as possible. I could not be more specific without knowing the language (or at least family of languages) used, and especially without knowing the architecture. If I were to assume a specific architectural context and ask a more specific question, a less experienced developer might have answered, "No, I don't have one of those" when in fact they had a logical equivalent for their ecosystem. This is a problem that also happens in the real world. Gaining context about the exact problem being solved is key when mentoring a less experienced programmer.''} (A5)
\end{quote}

\subsubsection{Unneeded or unrealizable situations} 
Some authors reported that some users' comments asked for a strategy to address situations or contexts that cannot occur and were not necessary. For example, one author stated: 
\begin{quote}
    \small 
    \textit{``It tells me that the reader is not a web developer and not familiar with web control objects like list-boxes (drop-down list boxes) and the term alpha for alphabetical. Web application[s] do not take symbols as input... as most user[s] are not aware of the extended ASCII set and more textboxes (or textareas) like the one I am entering this text in now, will not access the keyboard entry of Alt-214 for the alpha character.''} (A7)
\end{quote}
In other cases, users wanted a strategy to offer support for deciding a point in the strategy where there was not yet information available to make this decision.

\subsubsection{Mismatched level of knowledge}
Authors sometimes realized that they had misjudged the user's prior knowledge. The authors found it difficult to write a strategy without knowing the user's level of knowledge. Assuming background knowledge made some steps easier to understand or unnecessary to explain. Some authors stated that users should have been able to gain background knowledge through their investigation rather than through the strategy itself. One author responded:

\begin{quote}
    \small 
    \textit{``Debugging is not solely and purely related to CSS anymore, and it requires some debugging knowledge of JavaScript.''} (A8)

\end{quote}

\subsubsection{Hard-to-address feedback}
Some authors viewed some of the users' feedback to be hard to address. Some reported that some aspects of the strategy were hard to explain or would take too much time.  Other requests were impossible to satisfy, given the limitations of the strategy editing environment, such as requests for syntax highlighting or adding links to images.

\subsubsection{Resistance to increasing detail} 
Some users requested additional detail about specific strategy steps. Authors sometimes reported that adding these details would make their strategy too long and harder for users to follow, signaling the tension between detail and level of expertise. For example, one representative author said:

\begin{quote}
    \small 
    \textit{``It is hard to write [a] set of instructions so thoroughly to address all types of scenarios in simple words so everyone can understand, especially in the first try.''} (A8)
\end{quote}

\subsubsection{Ambiguous feedback} 
Some authors felt some comments left ambiguous exactly what the user requested. Authors often found broad requests for additional detail to be excessively vague. For example, one author wanted a more concrete example of the types of detailed information the users requested to include or the exact line number in the strategy where the user had gotten stuck. One author said:
\begin{quote}
    \small
    \textit{``I think not being specific and providing enough details in the comment about why the strategy action was ambiguous and didn't make much sense to them makes it hard to address. 
If the tester had specified why it doesn't make much sense, it could have been easier to address. ''} (A1)
\end{quote}
In other cases, users used a term or made a reference that the author did not understand. Some authors had difficulty understanding how the user was interpreting a statement. In cases where the authors found feedback ambiguous, some proposed including a screenshot of users' work to help them understand what they were doing and where they were getting stuck.

\subsection{Relationships Between Challenges Authoring and Using Strategies}

We hypothesized that many of the challenges that authors shared would have implications for the challenges that users of the strategies reported. To test this hypothesis, we analyzed the codes for the author's and users' challenges reported in Sections \ref{section:rq2} and \ref{section:rq3}. We created a file with all the reported challenges from authors accompanied by their corresponding strategy users' challenges. We paired each specific strategy's authoring difficulties with using difficulties. For each pair of difficulties, we counted the number of occurrences across all strategies. For each strategy, we counted at most one occurrence of each pair in cases where both users reported the challenge. Table \ref{table:challengesOccurances} lists the percent of paired occurrences of the total number of possible pairings. For example, for the \textit{'User perspective'} authoring difficulty and '\textit{Generality}' users' difficulty in Table \ref{table:challengesOccurances}, we counted the number of strategies that the pair of difficulties occurred. For the four out of five users who reported a generality difficulty, their authors reported having user perspective difficulty (80\%). We only reported the pairs that occurred in more than 30\% of the total cases in Table \ref{table:challengesOccurances}. The row 'Authors who reported the challenge' represents this number, which means out of 15 use cases, for each user challenge, we brought the challenge into the analysis if five or more users reported the challenge. The first column lists the fraction of authors who reported the challenge, out of a total of 19 authors.

\begin{table}[h]
\centering
\small
   \begin{tabular}{  l p{2.9cm}| lllll }
   
        \hline
        && \multicolumn{5}{c}{Users' Challenges}\\
        \cline{3-7}
        Authors' Challenges& 
        Authors who reported the challenge (\%) & 
        \rotatebox[origin=c]{90}{Generality}&
        \rotatebox[origin=c]{90}{Missing steps}&
        \rotatebox[origin=c]{90}{Imprecise steps}& \rotatebox[origin=c]{90}{Ambiguity}&
        \rotatebox[origin=c]{90}{\parbox{1.5cm}{Prerequisite  knowledge} }\\\hline
        \multicolumn{2}{c}{Users who reported the challenge (\%)}& 33.3& 66.7& 40& 46.7& 66.7\\\hline
        Generalization \& Abstract& 
        \centering 78.9& 80& 70& 100& 57& 70\\
        User Perspective& 
        \centering 63.2& 80& 60& 50& 42& 60 \\
        Level of details& 
        \centering 31.6& 40 &50 &16 & 14.3& 20 \\
        Completeness&      
        \centering 21.1& 40& 50& 50& 57.1& 50 \\
        Process& 
        \centering 26.3& 60& 40& 50& 14.3& 40 \\
        Effort& 
        \centering 78.9& 80& 70& 40& 46.7& 66.7\\\hline
   \end{tabular}

 \caption{The relationships between author and user challenges. For each challenge, the first row reports the percentage of users who reported the challenge, and the first column reports the percentage of users who reported the challenge. The remaining cells list the percentage of instances in which the challenges overlapped. }
\label{table:challengesOccurances}
\end{table}

       

As shown in Table \ref{table:challengesOccurances}, many, but not all, of the difficulties that users encountered were related to the difficulties that authors reported.
For example, 78.9\% of the authors reported difficulty making their strategy sufficiently general. They had difficulties adding an adequate amount of details and covering decent ranges of possible problem scenarios and edge cases while keeping the strategy simple. From this list, all of the users reported \textit{'Imprecise steps'} difficulty, 80\% reported having the \textit{Generality} using difficulty, and 70\% reported having \textit{Missing steps} and \textit{Prerequisite knowledge} challenges. For instance, A1 reported:

\begin{quote}
    \small
\textit{``strategy articulation is a bit hard because there may be multiple ways to troubleshoot and debug an issue, and some general strategy may not give an accurate way of diagnosing a problem. 
Also, a part of the strategy could be to ask a novice developer about the problem to come up with a proper strategy''} (Generalization, User perspective.) 
\end{quote}

U6, who used A1's strategy, reported 
\begin{quote}
    \small
\textit{``some explanation on how to debug edge cases could make it easier to follow, like how to debug animated elements or how to debug when element's style is manipulated with js. Also, debugging on inline styling could be helpful''} (Missing steps.)
\end{quote}

In another example, A5 reported:

\begin{quote}
    \small
\textit{``it's hard, but it's doable. The key is to put yourself in their shoes and keep reminding yourself of the audience. It also helps to visualize the novice developer and think about what's the first question he's going to ask when he sees this''} (User perspective). 
\end{quote}

U10, who used A5's strategy, reported the need for:

\begin{quote}
    \small
\textit{``more steps and more specific terms as to what errors can arise and why are they caused''}(Imprecise steps, Missing steps.)
\end{quote}





\section{Threats to Validity}
\label{sec:limitation}
As with any study, our study design had several vital threats to validity.

There were several potential threats to external validity in our design. First, our study differed from a work context in several ways. In encouraging participants to use the strategies to solve the problem, participants may not have benefited from the other resources they might normally use, such as asking teammate for help. Second, in the context of their own programming tasks, participants may face many different categories or even fewer challenges. Authoring strategies in a known context is also easier and less challenging; authors can communicate with their teammates about how one might need the strategy to be articulated.  Finally, in attempting to simulate the characteristics of a platform for sharing strategies that did not yet exist, the ways users and authors interacted might differ from how users might interact in a real-world platform. Users might experience difficulties that they did not experience in our study when working with longer or more challenging programming tasks. Platforms might incorporate different feedback mechanisms, such as multiple rounds of interactions between authors and users, or different ways of incorporating feedback. Our results are thus limited in partially reflecting characteristics of the specific platform we simulated.  

There were several potential threats to construct validity in our study design. First, because there are no widely accepted measures of prior knowledge in programming, the authors' expertise might have been more variable than intended. This might have led some developers to write strategies for tasks for which they had little expertise. Second, the ordinal survey questions, proposing possible difficulties, may add bias to the results by leading the authors only to focus on the proposed challenges and to forget to report other categories of challenges. Some of the challenges were reported only by one participant, and may not be broadly applicable.

From an internal validity perspective, our study design had several gaps. We did not directly observe users as they worked to examine how closely or carefully they followed the strategies. For example, one trade-off of not observing the users is that when a developer has better information than the strategy, a user might abandon performing the strategy's actions. So, a user might instead guess what could be difficult about using the strategy and critique it by comparing it to the strategy that they believe is more effective. 
In addition, authors might forget some of the challenges they faced after writing the strategy or mis-remembered them when reporting. 
When a user got stuck in a step in the strategy and needed to act in order to proceed, the user could give up continuing using the strategy. This would result in not collecting difficulties other users who progressed further might have experienced. In our design, authors were asked to author a strategy for a specific type of task, which users then performed. We did not examine any potential difficulties of identifying or choosing between relevant strategies for a specific task. Some participants were unfamiliar with the concept of a strategy, and we, therefore, provided training materials. While more experienced developers might not need this training, inexperienced developers might have benefited from further training, which might have reduced the difficulties they experienced.

\section{Discussion}
\label{sec:discussion}
In this paper, we examined the potential for experienced developers to share the strategies they use to solve programming problems. We investigated the strategies developers wrote as well as the challenges faced by both strategy authors and users. 
We found that experienced developers could explicitly articulate the strategies that existed in their heads, suggesting that it is possible for programming strategies to be externalized and shared. 

While experienced developers were able to write strategies down, our results revealed many challenges faced by authors in translating these into a form usable by other developers.
Our results identified five dimensions along which developers vary in their choices in how to articulate strategies. Future strategy sharing platforms should support authors in helping to consider these dimensions when authoring strategies, such as by providing feedback and suggestions from users along these dimensions and by offering reminders and explicit guidance. A sharing platform might, for example, include dimensions in a checklist, asking authors to consider describing how to set up referenced tools, check for irrelevant steps in the strategy, consider how many edge cases to cover, or suggest asking users to write down information for later use. Guidance might also be given to address difficulties in how to approach writing a strategy. In our study, many of the authors' challenges were related to not knowing their audiences, their needs, knowledge, and possible problem scenarios. Authors faced difficulties finding the right level of detail to include in the strategy, perhaps caused by the challenge of understanding what a less experienced developer needs. This difficulty may increase with more users of widely varying knowledge levels. A key challenge is designing mechanisms to align authored strategies with users' likely prior knowledge, possibly at authoring time, or when a developer is searching for strategies. 


Some challenges may be addressed through offering better tooling and representations for working with strategies, both in writing and using strategies.  Our findings suggest that authors found it hard to communicate some concepts through text, and might benefit from ways of expressing strategies with more visual content or even video demos. A strategy sharing platform might offer authors the ability to better integrate and connect strategies to reference tutorial materials or demos, which might, for example, illustrate through a video how to perform a specific step with a specific programming tool. Enhanced strategy editor support could also better support authors in writing valid Roboto syntax, through syntax highlighting and error messages.

When using strategies, strategy users were able to use the articulated strategies and provide feedback to the authors to improve strategies. Some of the users suggested alternative approaches to accomplish the strategy's goal to the authors to improve their strategies. Users' challenges were related to the level of knowledge necessary for using the strategy and the lack of some of that knowledge on the user side. However, our observations were limited in relying on developers' self-reported experiences, rather than direct observation. 


Over time, authors might improve their strategies to make them work for various problems at different scales, contexts, and languages, helping authors incrementally address challenges with abstraction, generalization, and scalability.
Iterative collaboration between authors and multiple users could help in covering more possible edge cases and addressing missing steps. In other situations, where a specific strategy is not a good fit, such as when it is inefficient or inapplicable, feedback might help clarify when a strategy is an effective approach and invite new strategies to be created for situations in which it is not an effective fit. Feedback may also sometimes be contradictory, and platforms should help authors in choosing between conflicting feedback and helping them explain the rationale to all users for the approaches they have taken. 



In making use of feedback from strategy users, strategy authors were able to understand some of the feedback they received and envision ways that they could improve their strategy. For example, a user's comment about adding a missing step in the strategy helped in clarifying the strategy. In other cases, authors disagreed with the feedback. Authors disagreed with users' feedback for several reasons. In some cases, authors felt that users had misused their strategy or realized that the context that the strategy is expected to be used in differed from the context in which the user was trying to use the strategy. 
Authors also found that they sometimes misjudged the users' prior knowledge, requiring users to have knowledge they did not have. 
Some of this feedback may trigger revisions, such as clarifying its applicability and prerequisite knowledge. Some of these issues might not even need intervention from authors, as users might clarify through their own experiences what knowledge is required and then share this with others. This also suggests the importance of effective search, to help users identify strategies that both address their specific situation at hand including its context as well as their knowledge and expertise.
In other cases, disagreements may reflect more fundamental differences, as in a case where a user believes that steps are inefficient or have a different idea in how to approach a strategy. The suggests that the boundary between user and author might sometimes be more fluid. A platform might, for example, let the user first suggest revisions to the author to accept or decline, and otherwise lets users fork and rewrite their own versions of strategies. 





There are many important questions on how strategy users should begin interacting with strategies, such as by discovering an existing strategy or requesting a strategy. In requesting a strategy, it is important that the strategy effectively match the context and expertise of the user that requested it. To address this, a sharing platform might require information from the user, such as completing a checklist of knowledge in technologies, concepts, and skills or even require the use of knowledge assessments to more accurately assess levels of expertise. This information could also be incorporated into browsing or search interfaces, helping identify strategies at the right level of expertise for users.


\section{Conclusion}
\label{sec:conclusion}
In this paper, we explored the potential for experienced developers to author programming strategies reflecting their hard-earned expertise in tackling everyday programming problems. We found that experts can write down their programming strategies, suggesting that, at least for experienced developers, programming can be a self-regulated and highly conscious activity. However, we found that the process of authoring strategies is full of challenging questions about generalizability and prior knowledge, suggesting that strategy authoring is much less like sharing code and more like teaching. Our results suggest the importance of feedback, demonstrating the ways in which feedback might help authors in improving strategies. Future platforms for sharing programming strategies should explore ways to facilitate easy and constructive communication between strategy authors and users to help improve strategies and enable their use by developers of any expertise level.
If future work can address these authoring challenges, explicit programming strategies may greatly improve the effectiveness of software developers by not only developing better tools but also by helping developers create, share, find, and use effective strategies for the problems they face. 

\begin{acknowledgements}
The authors would like to thank Stephen Hull for his helpful contributions to the analysis of our results. We thank our study participants for their time. This work was supported in part by the National Science Foundation under grants CCF-1703734 and CCF-1703304.

\end{acknowledgements}


\bibliographystyle{spmpsci}
\bibliography{references}

\end{document}